%% file: manuscript.tex
\documentclass[aps,prd,nofootinbib,superscriptaddress,reprint]{revtex4-2}
\pdfoutput=1
\usepackage{amsmath,amssymb,bm,bbm,booktabs,multirow,graphicx}
\usepackage[colorlinks=true, pdfstartview=FitH, linkcolor=blue, citecolor=blue, urlcolor=blue]{hyperref}

\graphicspath{{./plots/}}

\DeclareMathOperator\Tr{Tr}
\DeclareMathOperator\Pf{Pf}
\DeclareMathOperator\rme{\mathrm{e}}

\def\ckakko#1{\left\{ #1 \right\}}
\def\kkakko#1{\left[ #1 \right]}

\newcommand{\iquad}{\hspace{-10pt}}
\renewcommand{\Re}{\mathrm{Re}}
\newcommand{\der}{\partial}
\newcommand{\SU}{\text{SU}}
\newcommand{\U}{\text{U}}
\newcommand{\1}{\mathbbm{1}}
\newcommand{\I}{\mathcal{I}}
\newcommand{\RR}{\mathbb{R}}
\newcommand{\CC}{\mathbb{C}}
\newcommand{\rmd}{\mathrm{d}}
\newcommand{\T}{\textsf{T}}
\newcommand{\F}{\text{F}}
\newcommand{\A}{\text{A}}
\newcommand{\z}{\textsf{z}}
\newcommand{\w}{\textsf{w}}
\newcommand{\q}{\mathbf{q}}
\newcommand{\x}{\mathbf{x}}
\renewcommand{\epsilon}{\varepsilon}
\newcommand\eps\varepsilon
\renewcommand\phi\varphi
\newcommand\nrep{N_\text{rep}}
\newcommand\ev[1]{\left\langle#1\right\rangle}
\newcommand\AId{AI$^\dagger$}
\newcommand\AIId{AII$^\dagger$}
\newcommand\ket[1]{|#1\rangle}
\newcommand\bra[1]{\langle#1|}
\newcommand\scp[2]{\langle#1|#2\rangle}
\newcommand\me[3]{\langle#1|#2|#3\rangle}

\begin{document}
\title{New universality classes of the non-Hermitian Dirac operator in QCD-like theories}

\author{Takuya Kanazawa}
\affiliation{Research and Development Group, Hitachi, Ltd., Kokubunji, Tokyo 185-8601, Japan}
\author{Tilo Wettig}
\affiliation{Department of Physics, University of Regensburg, 93040 Regensburg, Germany}
\allowdisplaybreaks

\begin{abstract}
  In non-Hermitian random matrix theory there are three universality classes for local spectral correlations: the Ginibre class and the nonstandard classes \AId\ and \AIId. We show that the continuum Dirac operator in two-color QCD coupled to a chiral \U(1) gauge field or an imaginary chiral chemical potential falls in class \AId\ (\AIId) for fermions in pseudoreal (real) representations of \SU(2). We introduce the corresponding chiral random matrix theories and verify our predictions in lattice simulations with staggered fermions, for which the correspondence between representation and universality class is reversed. Specifically, we compute the complex eigenvalue spacing ratios introduced recently. We also derive novel spectral sum rules.
\end{abstract}
\maketitle

\input{introduction}
\input{dirac}
\input{rmt}
\input{lattice}
\input{numerics}
\input{conclusions}
\input{appendix}

\bibliography{manuscript.bbl}
\end{document}

%% file: introduction.tex
\section{Introduction}

The profound relationship between random matrix theory (RMT) and natural sciences has been the subject of research over decades \cite{Guhr:1997ve,Akemann:2011csh}. The energy-level statistics of quantum systems obeying RMT are considered to be a signature of quantum chaos \cite{Haakebook}, and the three universality classes of RMT (GOE/GUE/GSE) \cite{Dyson:1962es}, called the Wigner-Dyson (WD) classes, provide an exhaustive description of energy-level repulsion in non-integrable quantum systems \cite{Mehtabook}.\footnote{RMT ensembles that interpolate between WD classes have also been discussed \cite{PANDEY1981110,Mehta:1983ns,Pandey:1982br}. It turns out that a constant nonzero breaking of antiunitary symmetries of GOE and GSE always leads to GUE statistics in the large matrix-size limit \cite{PANDEY1981110}.} 

In the classification of Hermitian RMT \cite{Zirnbauer:1996zz,Altland:1997zz} there are seven non-WD symmetry classes. The three of them that preserve chiral symmetry \emph{exactly} describe the distribution of near-zero eigenvalues of the Euclidean Dirac operator in gauge theories with spontaneous chiral symmetry breaking \cite{Shuryak:1992pi,Verbaarschot:1993pm,Verbaarschot:2000dy,Verbaarschot2009}. The characteristics of the seven non-WD classes are manifest only in the vicinity of the spectral origin. In the bulk of the spectrum, i.e., far away from both the origin and the spectral edge, their level statistics reduce to those of the WD classes since the chiral symmetry and the particle-hole symmetry that ensure the \emph{global} pairing $\lambda\leftrightarrow -\lambda$ of the spectrum have no effect on \emph{local} correlations in the bulk. Indeed, lattice simulations of QED and QCD have revealed agreement between non-chiral RMT and the spectral statistics of the Dirac operator in the bulk \cite{Pullirsch:1998ke,Berg:1998xv}. It is therefore necessary to distinguish ``symmetry classes'' from ``universality classes''\footnote{In mathematics, the universality of RMT refers to the invariance of the spectral statistics for general matrix potentials \cite{DeiftBook,Erdos2011,Tao2012a,Tao2012b}.} --- the correct statement is that for Hermitian RMT there are 10 symmetry classes and 3 universality classes in total. 

Recent years have witnessed a surge of interest in non-Hermitian quantum systems \cite{Moiseyev2011,Benderbook,Ashida:2020dkc}. The symmetry classification of non-Hermitian RMT has been completed \cite{Kawabata:2018gjv,Zhou_2019}. There are 38 distinct symmetry classes.\footnote{Earlier classifications~\cite{bernard2001classification,Magnea:2007yk} contained 43 classes. The arXiv version of \cite{bernard2001classification} has been replaced with a corrected version that contains 38 classes. Hence there is no contradiction between \cite{bernard2001classification} and \cite{Kawabata:2018gjv}.} 
Out of the 38 classes, 22 possess chiral symmetry, and 3 of those describe the spectral statistics of the Dirac operator with baryon chemical potential, as reviewed in \cite{Akemann:2007rf,Akemann2011,Kanazawabook}. They are summarized below.
\begin{center}
\begin{tabular}{cc@{\hspace*{4.3pt}}c}
\toprule 
Class & Matrix form & Matrix elements
\\\midrule 
chiral real Ginibre & \multirow{3}{*}{
$\displaystyle \begin{pmatrix}0 & A \\ B & 0\end{pmatrix}$
}& real
\\
chiral complex Ginibre & & complex 
\\
chiral symplectic Ginibre & & quaternion real
\\\bottomrule 
\end{tabular}
\end{center}

Despite the diversity of non-Hermitian RMT, only one universality class has been known until recently: the Ginibre universality class \cite{Ginibre:1965zz}. For real, complex and quaternion Ginibre ensembles, the short-range eigenvalue correlations are always determined by the Vandermonde determinant $|\Delta(\{z_k\})|^2=\prod_{i<j}|z_i-z_j|^2$ (see \cite{Akemann_2019} for a summary of results). In \cite{Pullirsch:1998wp,Markum:1999yr}, agreement was found between the bulk spectral statistics of the Dirac operator with baryon chemical potential and the Ginibre ensemble.

It has recently been pointed out \cite{jaiswal2019universality,Hamazaki_2020} that there are in fact \emph{three} universality classes for local level statistics of non-Hermitian RMT:
\begin{center}
\begin{tabular}{c@{\hspace{10mm}}c}
\toprule 
Class & Matrix
\\\midrule 
Ginibre & $X$ 
\\
AI$^\dagger$ & $X^\T = X$
\\
AII$^\dagger$ & $X^\T = \sigma_2 X \sigma_2$
\\\bottomrule 
\end{tabular}
\end{center}
Here, the naming scheme of \cite{Hamazaki_2020} was adopted, $\T$ denotes transposition, and $\sigma_2$ is the second Pauli matrix. These three ensembles exhibit distinct nearest-neighbor eigenvalue spacing distributions \cite{jaiswal2019universality,Hamazaki_2020}. Complex symmetric random matrices have received little attention from physicists for a long time, with the exception of \cite{SFT1999}. 

In view of the aforementioned developments it is natural to ask the following questions. First, is there any application of the 19 nonstandard non-Hermitian chiral RMT classes to high-energy physics? Second, can the non-Ginibre universal statistics be observed in the Dirac spectrum? We answer both questions in the affirmative. Our contributions in this paper are as follows. 
(1) We show that Dirac operators in two-color QCD and adjoint QCD coupled to a \emph{chiral} $\U(1)$ gauge field (including an \emph{imaginary} chiral chemical potential as a special case) have special transposition symmetries. 
(2) We introduce two novel chiral RMTs that share the above symmetries.
(3) Through extensive lattice simulations with staggered fermions and using the complex spacing ratio analysis introduced in \cite{SRP2020} we find strong numerical evidence that spectral correlations of complex Dirac eigenvalues in these theories match those of the non-Ginibre universality classes of non-Hermitian RMT. 
(4) We highlight the physical importance of the Dirac operator coupled to a constant chiral gauge field.

Our findings are in stark contrast to the Dirac spectrum at finite baryon chemical potential, which exhibits correlations in the Ginibre class. This work provides a new point of view on quantum chaos in gauge theories. 

This paper is organized as follows. In Sec.~\ref{sec:dirac} we derive transposition properties of the continuum Dirac operator with fermions in real and pseudoreal representations of \SU(2) in the presence of a chiral \U(1) gauge field. In Sec.~\ref{sec:rmt} we present the non-Hermitian random matrix ensembles corresponding to these symmetries and derive an integral representation for the fermionic partition function. In Sec.~\ref{sec:lattice} we define the staggered lattice Dirac operators to be used in the numerical simulations and derive their transposition symmetries. We also derive novel spectral sum rules for these operators. In Sec.~\ref{sec:numres} we present our numerical results, which verify the correspondence between university classes and fermion representations in \SU(2). We conclude in Sec.~\ref{sec:conclusions}. Three appendices are provided for a discussion of constant chiral gauge fields and for technical details.


%% file: dirac.tex
\section{Properties of the Dirac operator in the continuum}
\label{sec:dirac}

In the following, the Euclidean Dirac matrices are denoted by $\gamma_\nu$ ($\nu=1,\ldots,4$), and we have $\gamma_5=\gamma_1\gamma_2\gamma_3\gamma_4$.
The Dirac operator is denoted by $D$. The continuum Dirac operator in the massless limit satisfies $\{D,\gamma_5\}=0$, which is known as chiral symmetry.

Let us first consider quarks in the fundamental representation of the gauge group $\SU(2)$. 
The Euclidean Dirac operator coupled to a chiral $\U(1)$ gauge field is given by
\begin{align}
  D & = \gamma_\nu(\der_\nu - i A^a_\nu \tau_a - i \gamma_5 B_\nu)\,,
      \label{eq:Ddefn}
\end{align}
where $A_\nu^a$ and $B_\nu$ are the $\SU(2)$ and $\U(1)$ gauge fields, respectively, and the $\tau_a$ are the generators of $\SU(2)$, i.e., the Pauli matrices acting in color space.
The $\U(1)$ charge assignment should be such that gauge anomalies are absent. 
As is well known, in the absence of $B_\nu$, $D$ possesses the antiunitary symmetry \cite{Kogut:2000ek}
\begin{align}
  [iD, C \tau_2 K] = 0\,,
\end{align}
where $C=i \gamma_4 \gamma_2$ is the charge conjugation matrix and $K$ is the complex conjugation operator. The coupling to $B_\nu$ breaks this symmetry, but $D$ retains the transposition symmetry
\begin{align}
  D^\T = C \tau_2 D C \tau_2\,. \label{eq:fusym}
\end{align}
As $C\tau_2$ is a symmetric unitary matrix, Theorem~3 in \cite{VERMEER2008382} can be applied, and it follows that $D=D^\T$ in a suitable basis. It is straightforward to verify that the argument so far holds for an arbitrary \emph{pseudoreal} representation of the gauge group.

Next, we turn to quarks in a real (e.g., adjoint) representation of a non-Abelian compact gauge group. In the absence of $B_\nu$, $D$ has the antiunitary symmetry \cite{Kogut:2000ek}
\begin{align}
  [iD,CK] = 0\,.
\end{align}
With the coupling to $B_\nu$, the antiunitary symmetry is broken, but $D$ fulfills the transposition symmetry
\begin{align}
  D^\T = CDC \,.
\end{align}
(In a real representation, the generators can always be chosen to be antisymmetric, see, e.g., \cite[Eq.~(5)]{Damgaard:2001fg}.)
Hua's decomposition \cite[Theorem~7]{Hua1944} implies that one can find a unitary matrix $U$ such that $C=U\Sigma_2U^\T$ with
\begin{align}
  \Sigma_2=\sigma_2\oplus\sigma_2\,,
\end{align}
where $\sigma_2$ acts on the Dirac indices. Performing a similarity transformation $D\to U^*DU^\T$ and using $C=-C^*$ it follows that $D$ can be cast into a complex quaternion form satisfying $D^\T=\Sigma_2 D \Sigma_2$. The following table gives a summary.
\begin{align}
  \begin{tabular}{c@{\hspace*{7mm}}l}
    \toprule
    Representation & \multicolumn{1}{c}{Symmetry of $D$} \\
    \midrule
    pseudoreal & $\{D,\gamma_5\}=0$ and $D^\T=D$ \\
    real & $\{D,\gamma_5\}=0$ and $D^\T=\Sigma_2D\Sigma_2$ \\
    \bottomrule
  \end{tabular}
  \label{tps}
\end{align}

Note that, if there is nonzero baryon chemical potential, the Dirac operator $D(\mu)\equiv D-\mu\gamma_4$ no longer satisfies \eqref{eq:fusym} since $D(\mu)^\T=C \tau_2 D(-\mu) C \tau_2$. In contrast, in the case of an \emph{isospin} chemical potential, $D_\text{I}(\mu)\equiv D-\mu\gamma_4 t_3$ satisfies $D_\text{I}(\mu)^\T=C \tau_2 t_2 D_\text{I}(\mu) C \tau_2 t_2$, where the $t_a$ are the generators of $\SU(2)$ flavor. Thus, by a rerun of the argument above, $D_\text{I}(\mu)^\T=\Sigma_2 D_\text{I}(\mu) \Sigma_2$ in a suitable basis. 

If we drop the spatial components of $B_\nu$ and make the temporal component constant, the latter represents an imaginary chiral chemical potential,
\begin{align}
  D & = \gamma_\nu (\der_\nu-iA_\nu^a \tau_a) + \mu_5\gamma_5\gamma_4\,,
      \quad \mu_5 \in i\,\RR\,. \label{eq:3sdf}
\end{align}
We comment on the physical implication of this term in App.~\ref{app:axial}. 
Recently \cite{Felski:2020vrm} has investigated the effect of a term analogous to \eqref{eq:3sdf} in the NJL model to formulate $\mathcal{PT}$-symmetric chiral symmetry breaking. 
We note in passing that a \emph{real} chiral chemical potential has been extensively used in lattice simulations to create chirality imbalance \cite{Yamamoto:2011gk,Braguta:2015zta,Braguta:2015owi}, see \cite{Yang:2020ykt} for a review. 


%% file: rmt.tex
\section{\boldmath Nonstandard chiral random matrix theory}
\label{sec:rmt}

\subsection{Definition of ensembles}
\label{sec:ensembles}

On the basis of the transposition symmetries \eqref{tps} we propose that the spectral properties of these Dirac operators are described by the following non-Hermitian random matrices:
\begin{center}
  \begin{tabular}{c@{\hspace*{4mm}}c@{\hspace*{4mm}}c}
    \toprule 
    Representation & Matrix form & Matrix elements
    \\\midrule
    pseudoreal & $\begin{pmatrix}0&V \\ V^\T & 0\end{pmatrix}$ & complex \vspace{2mm}\\
    real & $\begin{pmatrix}0&V \\ \sigma_2 V^\T \sigma_2 & 0\end{pmatrix}$ & complex quaternion
    \\\bottomrule
  \end{tabular}
\end{center}
The matrix elements of $V$ obey independent Gaussian distributions with the same variance and mean zero. 
The ensemble in the first row corresponds to class \AId\ with sublattice symmetry $\mathcal{S}_+$ (equivalent to class D with sublattice symmetry $\mathcal{S}_+$), and the ensemble in the second row to class \AIId\ with sublattice symmetry $\mathcal{S}_+$ (equivalent to class C with sublattice symmetry $\mathcal{S}_+$), respectively \cite[Tables VII, XII and XIII]{Kawabata:2018gjv}. Note that if $V$ is real or quaternion real, these ensembles reduce to chiral GOE and chiral GSE, respectively. Every eigenvalue of the ensemble in the second row is doubly degenerate due to the non-Hermitian generalization of Kramers' theorem \cite{Kawabata:2018gjv}. 

In section~\ref{sec:numres} we perform extensive numerical analyses to verify our proposal. In this paper, we focus on \emph{bulk} properties of the spectrum and leave the analysis of spectral statistics near the origin to future work.

\subsection{Fermionic partition function}

The fermionic partition function, i.e., the average of the product of characteristic polynomials, is of special interest in RMT. The arguments for pseudoreal and real quarks can be run in parallel, so we shall concentrate on pseudoreal quarks in the following. At first sight it may seem natural to consider the partition function
\begin{align}
  Z = \int_{\CC^{N\times N}} \iquad 
  \rmd V\;\rme^{-N \Tr VV^\dagger} {\det}^{N_f}
  \begin{pmatrix}m & V \\ V^\T & m \end{pmatrix},
\end{align}
where $N_f$ is the number of quark flavors of mass $m$,
but this $Z$ is pathological as it vanishes in the chiral limit $m\to0$. Instead we must consider $K$ pairs of conjugate quarks \cite{Stephanov:1996ki}, 
\begin{align}
  Z_K & = \int_{\CC^{N\times N}} \iquad \rmd V~
	\rme^{-N \Tr VV^\dagger}
	\prod_{f=1}^{K}\left|\det \begin{pmatrix}z_f &V\\V^\T&w_f\end{pmatrix} \right|^2,
\end{align}
where $w_f$ and $z_f$ are masses. We consider even $N$. 
The determinants may be expressed as a Grassmann integral,
\begin{align}
  Z_K
  & \propto \int_{\CC^{N\times N}} \iquad 
    \rmd V ~\rme^{-N \Tr VV^\dagger}
    \int \rmd(u, \bar{u}, d, \bar{d})
  \notag \\
  & \quad \times \exp\bigg[
    \begin{pmatrix}\bar{u}_{Lf} \\ \bar{u}_{Rf} \end{pmatrix}^\T 
    \begin{pmatrix}z_f&V\\V^\T&w_f\end{pmatrix}
    \begin{pmatrix}u_{Rf} \\ u_{Lf} \end{pmatrix} 
  \notag \\
  & \qquad\qquad + \begin{pmatrix}\bar{d}_{Lf} \\ \bar{d}_{Rf} \end{pmatrix}^\T 
  \begin{pmatrix}z_f^*&V^*\\V^\dagger&w_f^*\end{pmatrix}
  \begin{pmatrix}d_{Rf} \\ d_{Lf} \end{pmatrix}
  \bigg]\,.
\end{align}
After integrating out $V$ we introduce auxiliary bosonic fields
\begin{equation}
  \begin{alignedat}{2}
    C_L &\sim \bar{u}_L \bar{d}_L\,,\quad
    & C_R &\sim \bar{u}_R \bar{d}_R\,, \\
    D_L &\sim \bar{d}_R u_L\,,
    & D_R &\sim \bar{d}_L u_R
  \end{alignedat}
\end{equation}
to perform a Hubbard-Stratonovich transformation. The Grassmann variables can then be integrated out trivially to yield
\begin{align}
  Z_K &\propto \int_{\CC^{2K\times 2K}} \iquad \rmd \Omega~
	\rme^{-N \Tr \Omega \Omega^\dagger}
	{\det}^{N/2}\begin{pmatrix} \Omega &\w \\ \w^*& \Omega^\T \end{pmatrix}
	\notag \\
      & \quad \times {\det}^{N/2}\begin{pmatrix} \Omega^*&\z \\ \z^*&\Omega^\dagger \end{pmatrix},
   \label{eq:dsf4}
\end{align}
where
\begin{align}
  \Omega \equiv \begin{pmatrix}C_L & -D_L^* \\ D_R & C_R^* \end{pmatrix}, \quad
  \w \equiv \begin{pmatrix}0&-iw\\iw&0\end{pmatrix}, \quad
  \z \equiv \begin{pmatrix}0&-iz\\iz&0\end{pmatrix}
\end{align}
with $z=\text{diag}(z_f)$ and $w=\text{diag}(w_f)$. 
In the microscopic large-$N$ limit with $\w\sim \z \sim 1/\sqrt{N}$, the integrand of \eqref{eq:dsf4} is $\bigl[\rme^{-\Tr \Omega\Omega^\dagger}\det(\Omega \Omega^\dagger)\bigr]^N$. After a singular-value decomposition the saddle point is given by $\Omega=\1_{2K}$. This implies spontaneous symmetry breaking $\U(2K)\times\U(2K) \to \U(2K)_\text{diag}$ \cite{Shuryak:1992pi}. The soft mode around the saddle point may be parametrized as $\Omega=U\in\U(2K)$. Then
\begin{align}
  \iquad 
  Z_K & \propto \int_{\U(2K)}\iquad \rmd U~ {\det}^{N/2}\bigg[
        \begin{pmatrix} U & \w \\ \w^*& U^\T \end{pmatrix}
        \begin{pmatrix} U^\dagger & \z^* \\ \z & U^* \end{pmatrix}\bigg]
        \notag \\
      & \propto \int_{\U(2K)}\iquad \rmd U~\exp \bigg[
	\frac{N}{2}\Tr(U\z^*U^\T \z + \w U^* \w^* U^\dagger) \bigg].
        \label{eq:ZK}
\end{align}
This is the most generic result in the large-$N$ microscopic limit. 

Let us consider the special case $\forall w_f=\forall z_f= \lambda/\sqrt{N}$. With the skew-symmetric matrix $\I\equiv\begin{pmatrix}0&-\1_{K}\\\1_{K}&0\end{pmatrix}$ we have
\begin{align}
  Z_K \propto \int_{\U(2K)}\iquad \rmd U~\exp \big[ |\lambda|^2 
	\Re\Tr(U\I U^\T \I)\big]\,.
\end{align}
This integral has been evaluated analytically in \cite{Smilga:1994tb}, which enables us to obtain
\begin{align}
  Z_K & \propto \underset{1\leq i, j \leq 2K}{\Pf}
	\left[(j-i)I_{i+j-2K-1}(2|\lambda|^2)\right],
\end{align}
where $\Pf$ denotes the Pfaffian and $I$ is the modified Bessel function of the first kind.


%% file: lattice.tex
\section{Lattice Dirac operators}
\label{sec:lattice}

In this section we define the lattice Dirac operators that will be used in the numerical simulations. To keep the notation simple we will use the same symbol $D$ as for the continuum Dirac operator since confusion is unlikely to arise. We work in four dimensions. All dimensionful quantities are given in lattice units, i.e., we set the lattice spacing $a$ to 1.

\subsection{Staggered Dirac operator with chiral \U(1) field}

The massless staggered Dirac operator coupled to a lattice gauge field $U_\mu(x)$ is given by its matrix elements
\begin{align}
  \label{eq:stagg}
  D_{xy} & = \frac{1}{2} \sum_{\mu=1}^{4}\eta_\mu(x) \kkakko{U_\mu(x)\delta_{x+\mu,y} - U_\mu(y)^\dagger \delta_{x,y+\mu}},
\end{align}
where $\mu$ is a unit vector in one of the four dimensions and the $\eta_\mu(x)$ are the usual staggered phases. We consider $\SU(2)$ gauge fields but do not specify the representation for the time being.

The continuum chiral symmetry is replaced by the remnant chiral symmetry $\{D,\eps\}=0$, where the operator $\eps$ is given by its matrix elements $\eps_{xy}=\eps(x)\delta_{xy}$ with
\begin{align}
  \eps(x)=(-1)^{x_1+x_2+x_3+x_4}\,.
\end{align}
Since $D$ is anti-Hermitian its eigenvalues are purely imaginary. It is straightforward to show that due to $\{D,\eps\}=0$ the eigenvalues of $D$ come in pairs $\pm\lambda$ with eigenvectors $\psi$ and $\eps\psi$. 

We now introduce a chiral $\U(1)$ gauge field
\begin{align}
  \theta_\mu(x)=\exp(i\eps(x)\phi_\mu(x))\,,  
\end{align}
where $\phi_\mu(x)$ is a smooth real gauge field. The staggered Dirac operator coupled to $\theta_\mu(x)$ reads
\begin{align}
  \label{eq:U1}
  D(\theta)_{xy} = \frac{1}{2} \sum_{\mu=1}^{4}\eta_\mu(x)
  & \bigl[U_\mu(x)\theta_\mu(x)\delta_{x+\mu,y} \notag\\
  &- U_\mu(y)^\dagger\theta_\mu(y)\delta_{x,y+\mu}\bigr]\,.
\end{align}
The $\U(1)$ gauge transformations of the fermions and the gauge field are given by
\begin{subequations}
  \begin{align}
    \chi(x) & \to \rme^{i\epsilon(x)\alpha(x)}\chi(x)\,,\\
    \bar\chi(x) & \to \bar\chi(x) \rme^{i\epsilon(x)\alpha(x)}\,,\\
    \varphi_\mu(x) &\to \varphi_\mu(x)-\alpha(x)+\alpha(x+\mu)\,,
  \end{align}
\end{subequations}
where $\alpha(x)$ is an arbitrary gauge potential.

For $\theta_\mu(x)\ne1$, $D(\theta)$ is no longer anti-Hermitian so that its eigenvalues move from the imaginary axis into the complex plane. The remnant chiral symmetry $\{D(\theta),\eps\}=0$ still holds, and therefore the complex eigenvalues still come in pairs $\pm\lambda$ with eigenvectors $\psi$ and $\eps\psi$.

We now consider the $\SU(2)$ gauge field in the fundamental and adjoint representation, denoted by $U^\F$ and $U^\A$, respectively. The latter is obtained from the former by
\begin{align}
  \label{eq:f2a}
  U_\mu^\A(x)_{ab}=\frac12\Tr(\tau_aU_\mu^\F(x)\tau_bU_\mu^\F(x)^\dagger)\,.
\end{align}
It is straightforward to check that $U^\A$ is real.

All statements made in this section so far hold for both representations. The difference between the representations lies in the transposition symmetries. Using the pseudoreality $(U^\F)^*=\tau_2 U^\F \tau_2$ it is straightforward to show that
\begin{align}
  D^\F(\theta)^\T = - \tau_2 D^\F(\theta) \tau_2\,.
	\label{eq:2453ds}
\end{align}
As a consequence, the eigenvalues of $D^\F(\theta)$ are twofold degenerate. We show this in App.~\ref{app:twofold}.
Similarly, for the adjoint representation we use $(U^\A)^*=U^\A$ to show that
\begin{align}
  D^\A(\theta)^\T = - D^\A(\theta)\,.
	\label{eq:fds425}
\end{align}
We therefore expect $D(\theta)$ in the fundamental or adjoint representation of $\SU(2)$ to be in universality class \AIId\ or \AId, respectively.\footnote{The minus signs in \eqref{eq:2453ds} and \eqref{eq:fds425} compared to \eqref{tps} correspond to an additional factor of $-1$ in one of the off-diagonal blocks in the table of Sec.~\ref{sec:ensembles}. This leads to a relative factor of $i$ in the eigenvalues, which has no effect on the bulk spectral correlations.} This expectation will be confirmed numerically in Sec.~\ref{sec:numres}. We note that the universality classes of the fundamental and adjoint representation are interchanged for staggered fermions compared to the continuum. This is a generic phenomenon \cite{Halasz:1995vd,BerbenniBitsch:1997tx,Edwards:1999px}. 

The eigenvalues of $D(\theta)$ satisfy a sum rule that is very useful as a check of the eigenvalues obtained numerically. We show in App.~\ref{app:sumrules} that
\begin{align}
  \label{eq:sumrule_U1}
  \Tr D(\theta)^2=\sum_n\lambda_n^2=-2\nrep V\big\langle\theta_\mu^2(x)\big\rangle_{x\mu}\,,
\end{align}
where the average is over all links of the lattice, $V$ is the lattice volume and $\nrep$ is the dimension of the representation, i.e., for gauge group SU(2) we have $\nrep=2$ ($\nrep=3$) in the fundamental (adjoint) representation, respectively.

\subsection{Staggered Dirac operator with chiral chemical potential}
\label{sec:mu5}

To introduce the chiral potential $\mu_5$ we follow Braguta et al.~\cite{Braguta:2015zta} but slightly change their notation to conform with our conventions. We write the staggered Dirac operator in the presence of $\mu_5$ in the form
\begin{align}
  \label{eq:mu5}
  D(\mu_5)=D+\frac12\mu_5D_5\,,
\end{align}
where $D$ is the operator in \eqref{eq:stagg} and $D_5$ is given by its matrix elements
\begin{align}\label{eq:D5}
  (D_5)_{xy}&=s(x)\left[\bar U_\delta(x)\delta_{x+\delta,y}+\bar U_\delta^\dagger(y)\delta_{x,y+\delta}\right]
\end{align}
with $s(x)=(-1)^{x_2}$, $\delta=(1,1,1,0)$ and 
\begin{align}\label{eq:Udelta}
  \bar U_\delta(x)=\frac16\sum_{i,j,k=\text{perm}(1,2,3)}U_i(x)U_j(x+\hat i)U_k(x+\hat i+\hat j)\,.
\end{align}
In the continuum limit, the second term in \eqref{eq:mu5} reduces to the $\mu_5$ term in \eqref{eq:3sdf} \cite{Braguta:2015zta}. It is straightforward to show that $D_5$ is anti-Hermitian and anticommutes with $\eps$. Thus, for $\mu_5\in\RR$ the eigenvalues of $D(\mu_5)$ are purely imaginary, while for $\mu_5\notin\RR$ they are generically complex. Furthermore, we still have $\{D(\mu_5),\eps\}=0$, and thus the eigenvalues of $D(\mu_5)$ again come in pairs $\pm\lambda$ with eigenvectors $\psi$ and $\eps\psi$. Equations~\eqref{eq:2453ds} and \eqref{eq:fds425} also hold with $D(\theta)$ replaced by $D(\mu_5)$, and hence we expect the same correspondence between representation and universality class. For $\SU(2)$ fundamental every eigenvalue of $D(\mu_5)$ is again twofold degenerate, see App.~\ref{app:twofold}.

The eigenvalues of $D(\mu_5)$ also satisfy a useful sum rule. We show in App.~\ref{app:sumrules} that
\begin{align}
  \label{eq:sumrule_mu5}
  \Tr D(\mu_5)^2=-V\Bigl[2\nrep+\frac12\mu_5^2\bigl\langle\Tr\bar U_\delta(x)\bar U_\delta^\dagger(x)\bigr\rangle_x\Bigr],
\end{align}
where the average is over all sites of the lattice.


%% file: numerics.tex
\section{Numerical results}\label{sec:numres}

The aim of this section is to test our proposal by numerical simulations of the corresponding lattice quantum field theories. As mentioned above, we use staggered fermions. We work in the quenched approximation since it is numerically cheap and sufficient for the point we are making. The inclusion of dynamical fermions does not change the universality class and therefore makes no difference for the quantities we compute.

\subsection{Overview of the simulations}

\begin{figure}
  \centering
  \begin{tabular}{cc}
    \includegraphics[width=.484\columnwidth]{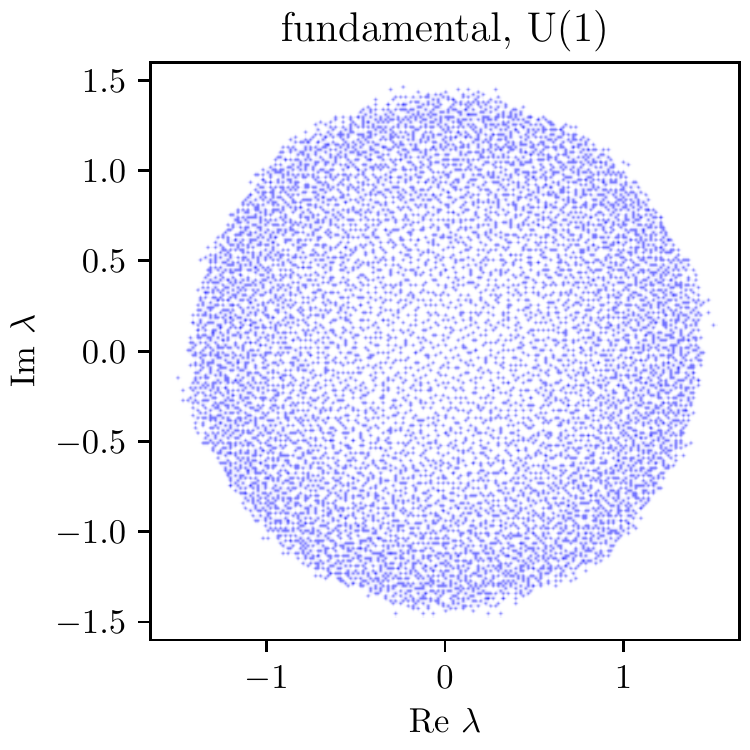}
    & \includegraphics[width=.484\columnwidth]{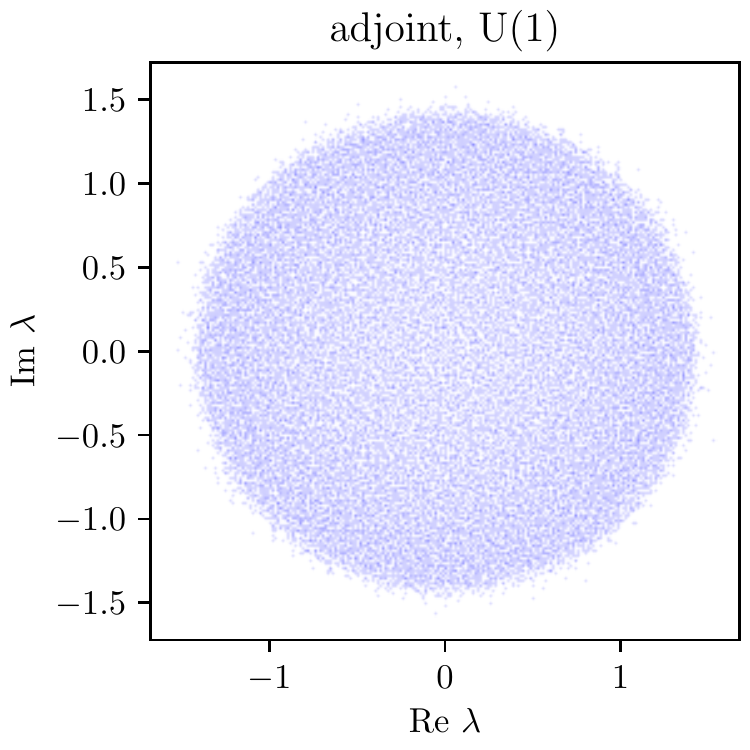} \\
    \includegraphics[width=.484\columnwidth]{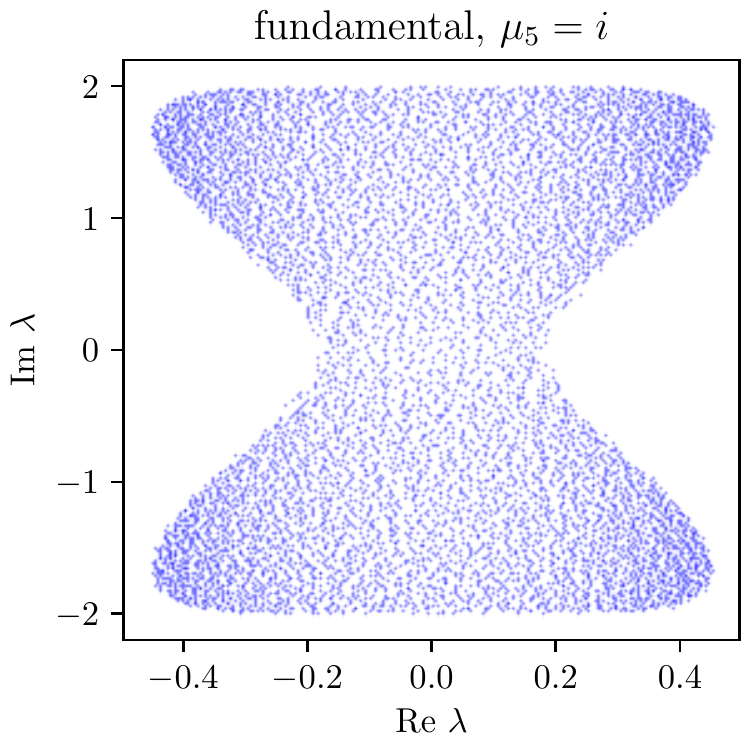}
    & \includegraphics[width=.484\columnwidth]{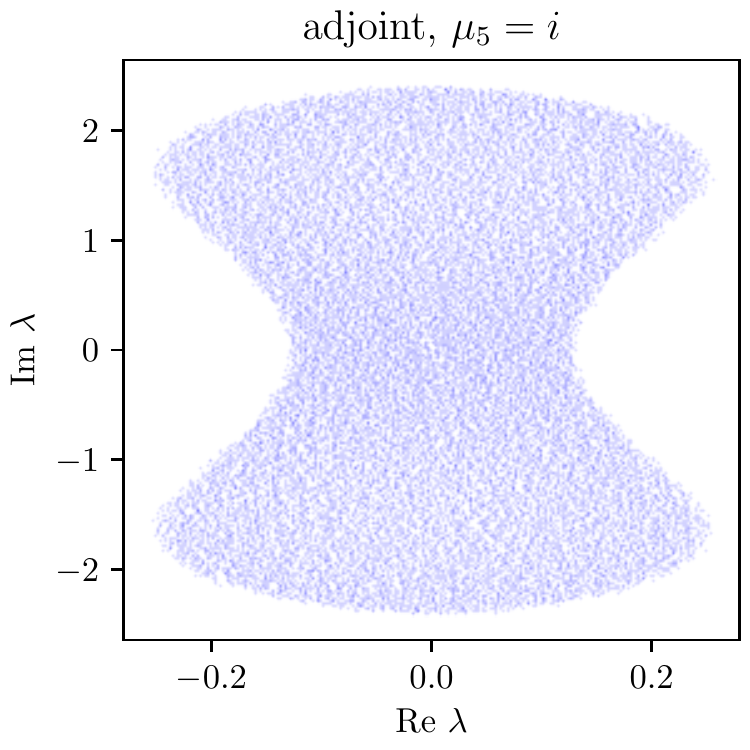} \\
    \includegraphics[width=.484\columnwidth]{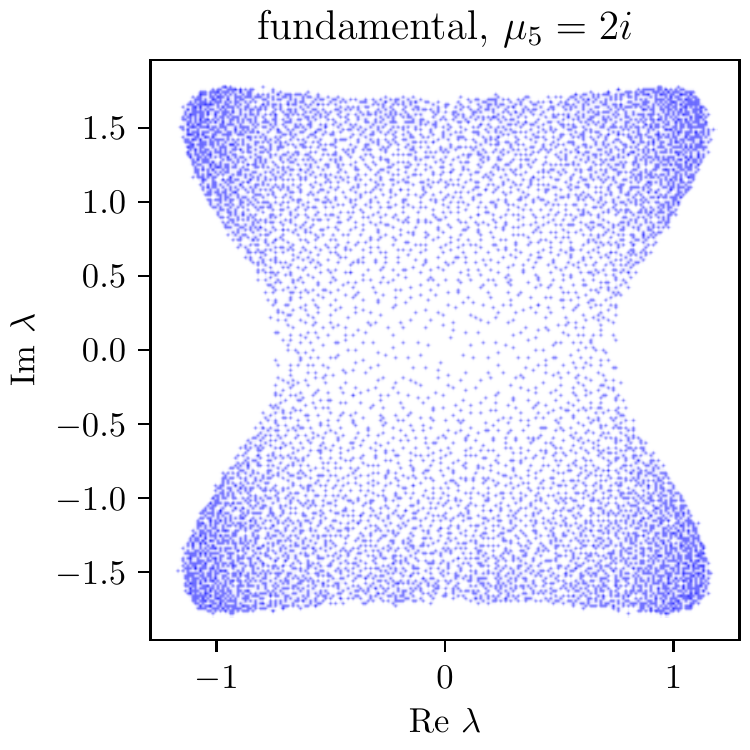}
    & \includegraphics[width=.484\columnwidth]{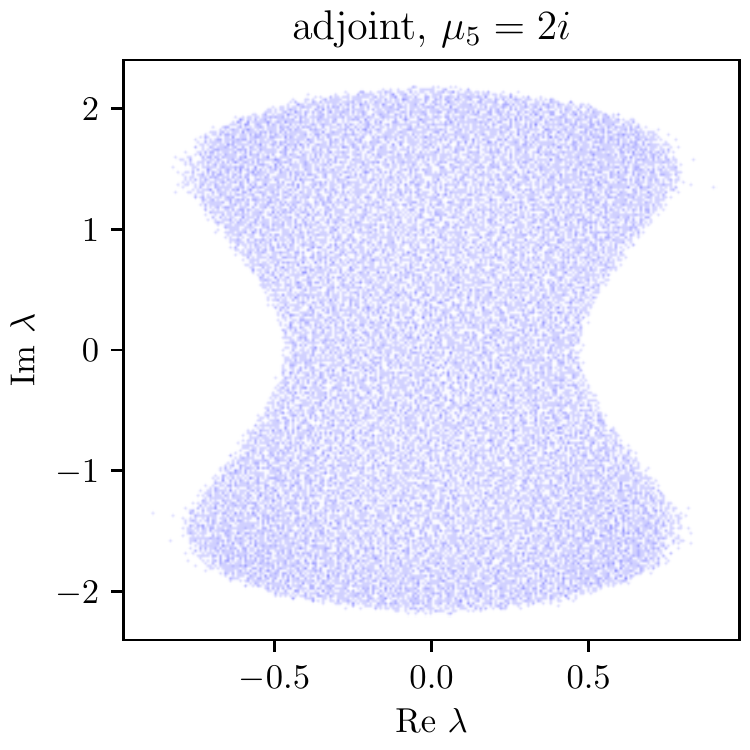}
  \end{tabular}
  \caption{Scatter plots of the staggered Dirac eigenvalues for a single configuration in SU(2) lattice gauge theory on an $8^3\times16$ lattice. Left: fermions in the fundamental representation of SU(2), right: fermions in the adjoint representation. In the top row, a chiral U(1) field is included. In the other two rows, an imaginary chiral potential $\mu_5$ is included.}
  \label{fig:scatter}
\end{figure}

Our simulations were done in the gpt framework \cite{gpt}, which provides a convenient python interface to the Grid library \cite{Boyle:2015tjk}. We used the simple (unimproved) Wilson plaquette action for the gauge fields. The $\SU(2)$ gauge field in the fundamental representation was generated using the Creutz-Kennedy-Pendleton heatbath algorithm \cite{Creutz:1980zw, Kennedy:1985nu}. The adjoint $\SU(2)$ field was obtained from the fundamental field via \eqref{eq:f2a}.
For the generation of the compact U(1) field $\exp(i\phi_\mu(x))$ we used the Hattori-Nakajima heatbath algorithm \cite{Hattori:1992qk}. The coupling constants were chosen as $\beta_\text{SU(2)}=4/g^2=2.0$ and $\beta_\text{U(1)}=1/e^2=0.9$, which in both cases corresponds to the confined phase.  

Numerically, the most expensive operation is the calculation of all eigenvalues of the Dirac operator, which scales like $O(V^3)$. Fortunately, for the comparison of lattice data and RMT predictions, a relatively small volume is sufficient since the local spectral correlations have very small finite-volume effects. We found that a lattice size of $8^3\times16$ already shows nearly perfect agreement with RMT and therefore restrict ourselves to this lattice size. In this case the eigenvalue calculation can be done on a single CPU in a very short time, i.e., we do not need to employ HPC resources. The number of configurations for each case is given in Table~\ref{tab:configs}.

\begin{table}
  \centering
  \begin{tabular}{l@{\hspace{.05\columnwidth}}|@{\hspace{.05\columnwidth}}c@{\hspace{.05\columnwidth}}c@{\hspace{.05\columnwidth}}c}
    & U(1) & $\mu_5=i$ & $\mu_5=2i$ \\\hline
    SU(2) fund. & 271 & 415 & 415 \\
    SU(2) adjoint & 267 & 267 & 266
  \end{tabular}
  \caption{Number of configurations for the cases simulated. $\U(1)$ refers to the operator in \eqref{eq:U1}, while the last two columns give the value of $\mu_5$ in \eqref{eq:mu5}.}\vspace*{-1mm}
  \label{tab:configs}
\end{table}

\begin{figure}[b]
  \centering
  \begin{tabular}{cc}
    \includegraphics[width=.484\columnwidth]{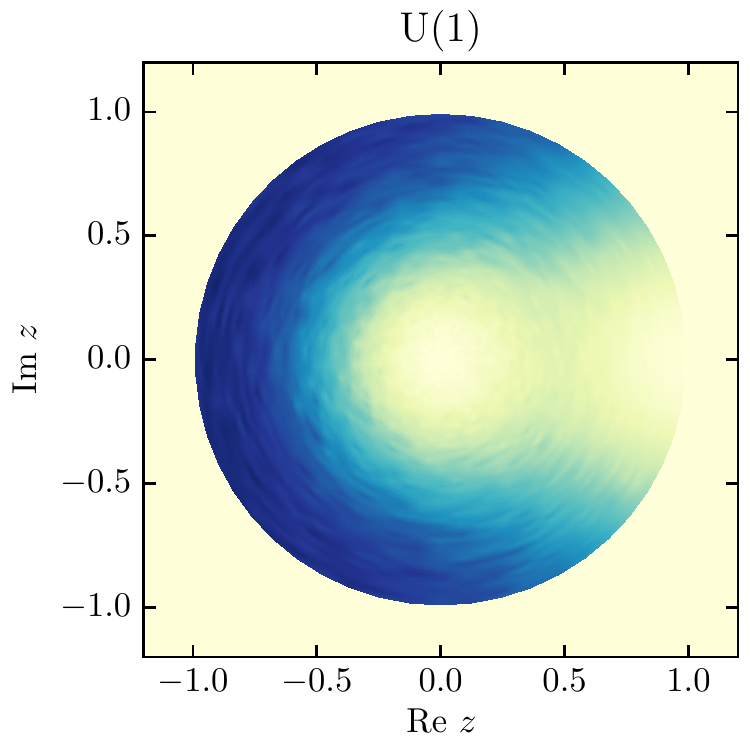}
    & \includegraphics[width=.484\columnwidth]{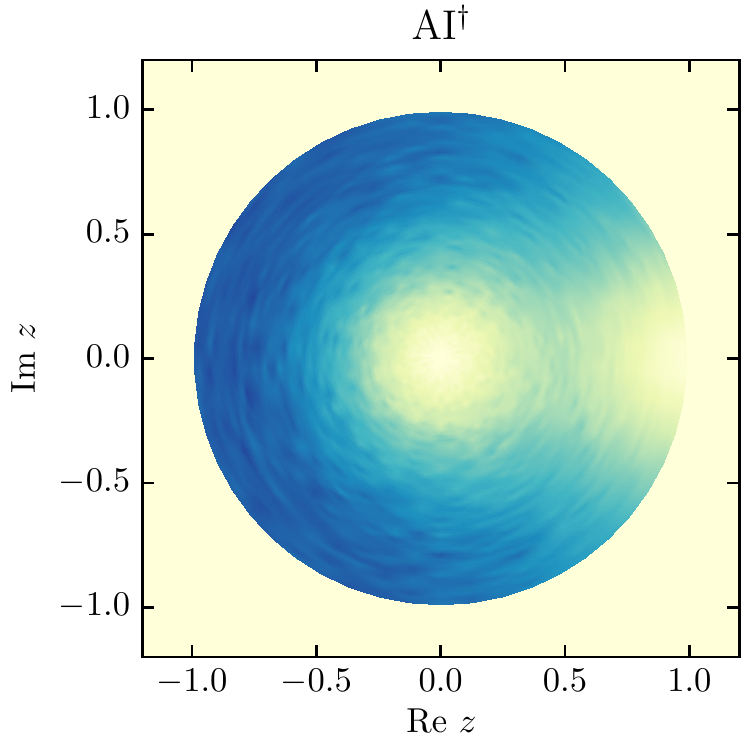} \\
    \includegraphics[width=.484\columnwidth]{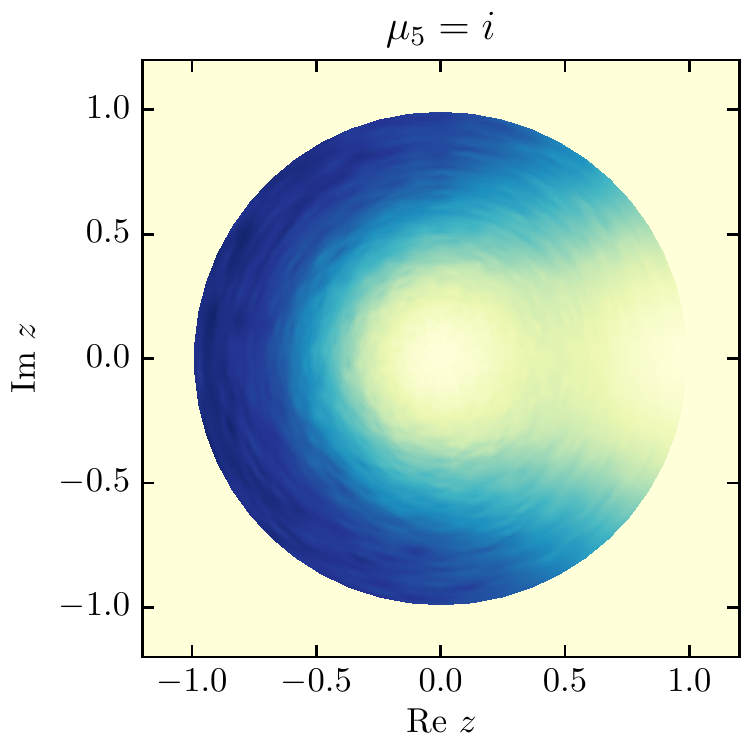}
    & \includegraphics[width=.484\columnwidth]{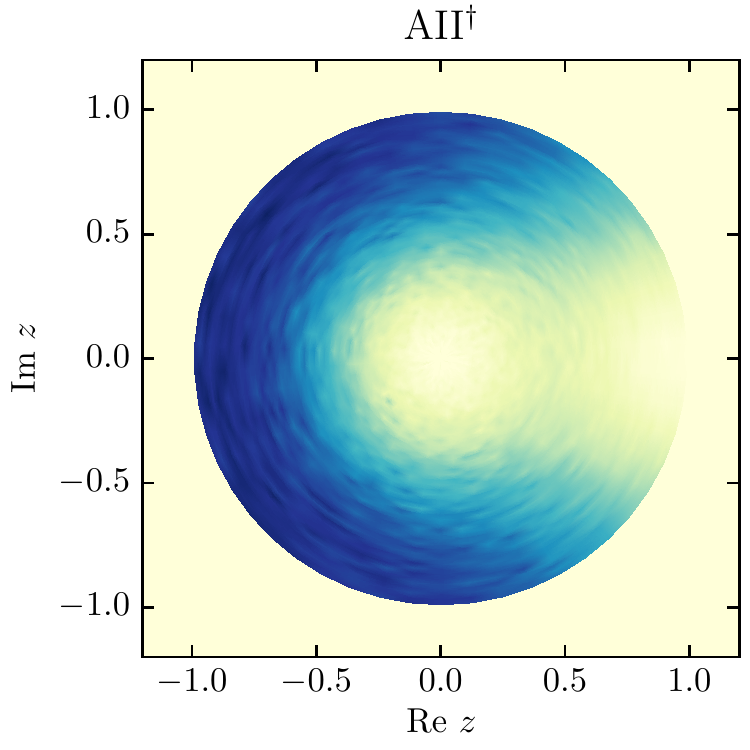} \\
    \includegraphics[width=.484\columnwidth]{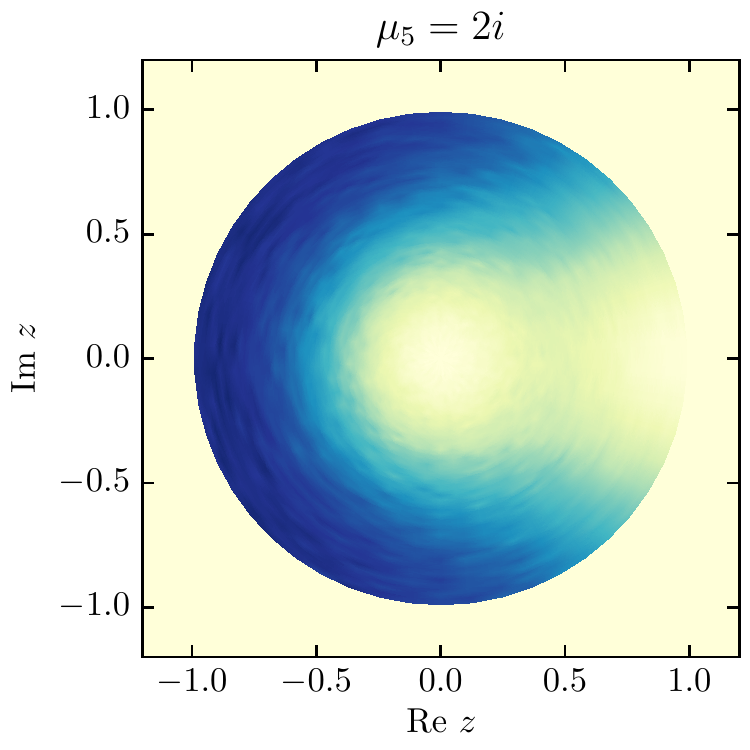} 
    & \includegraphics[width=.484\columnwidth]{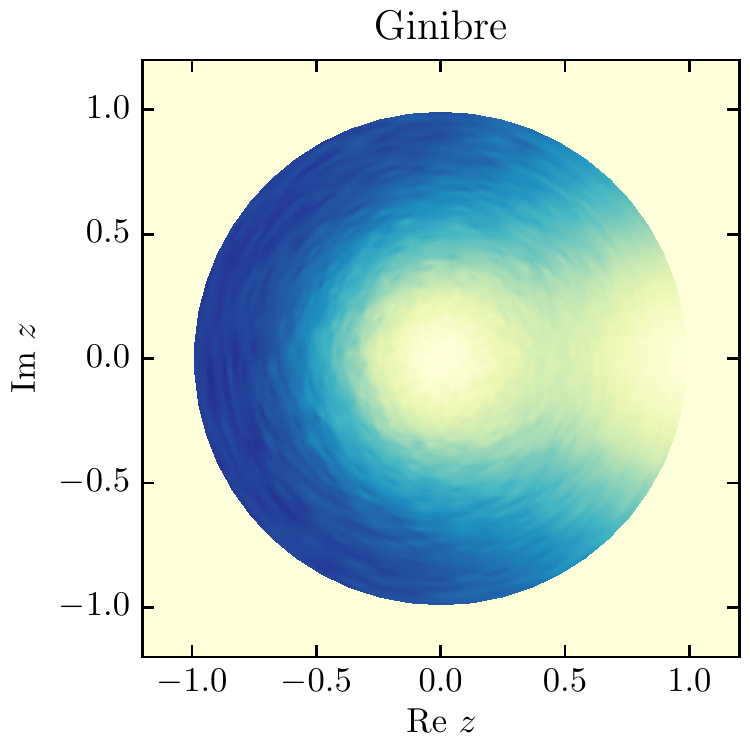}
  \end{tabular}
  \caption{Heatmaps of $P(z)$ for staggered SU(2) lattice data in the fundamental representation with chiral U(1) field or imaginary chiral chemical potential (left) and for three RMT ensembles (right).}
  \label{fig:2d_fund}
\end{figure}

Scatter plots of the Dirac eigenvalues for a single configuration are shown in Fig.~\ref{fig:scatter} for the cases we considered. The spectral correlations on the scale of the mean level spacing are insensitive to the global features of the spectrum. The main purpose of Fig.~\ref{fig:scatter} is to demonstrate that the distribution of the eigenvalues is smooth, i.e., there are no clustering effects. Note that for a fixed lattice volume the number of independent eigenvalues for the adjoint representation is larger by a factor of 3 compared to the fundamental representation since (i) the dimension of the representation is larger by a factor of 3/2 and (ii) the eigenvalues in the fundamental representation are twofold degenerate.

Once the eigenvalues $\lambda_k$ have been obtained, we follow \cite{SRP2020} and compute the complex spacing ratios
\begin{align}
  \label{eq:zk}
  z_k=\frac{\lambda_k^\text{NN}-\lambda_k}{\lambda_k^\text{NNN}-\lambda_k}\,,
\end{align}
where $\lambda_k^\text{NN}$ and $\lambda_k^\text{NNN}$ are the nearest and next-to-nearest neighbor of $\lambda_k$ in the complex plane. We used a $k$-d tree algorithm to perform the nearest-neighbor search. We included all eigenvalues in the computation of the $z_k$. This causes some boundary effects, see the discussion in \cite{SRP2020}, but these effects are suppressed in large volumes. 

Writing $z=r\rme^{i\theta}$ we have $r\le1$ by construction.\footnote{We use the same symbol $\theta$ for the phase angle of $r$ as for the chiral $\U(1)$ field. Since the two quantities never appear in the same section we hope that no confusion arises.} We computed the distributions $P(z)$, $P(r)$ and $P(\theta)$ as well as moments of these distributions from the lattice data and compared them with the corresponding RMT predictions.
To the best of our knowledge, analytical RMT results for these quantities are not available yet. We therefore generated the RMT predictions numerically by diagonalizing 300 random matrices of dimension 4000 for every ensemble.

\subsection{Results for SU(2) fundamental}
\label{sec:fund}

\begin{figure}
  \centering
  \begin{tabular}{cc}
    \includegraphics[width=.484\columnwidth]{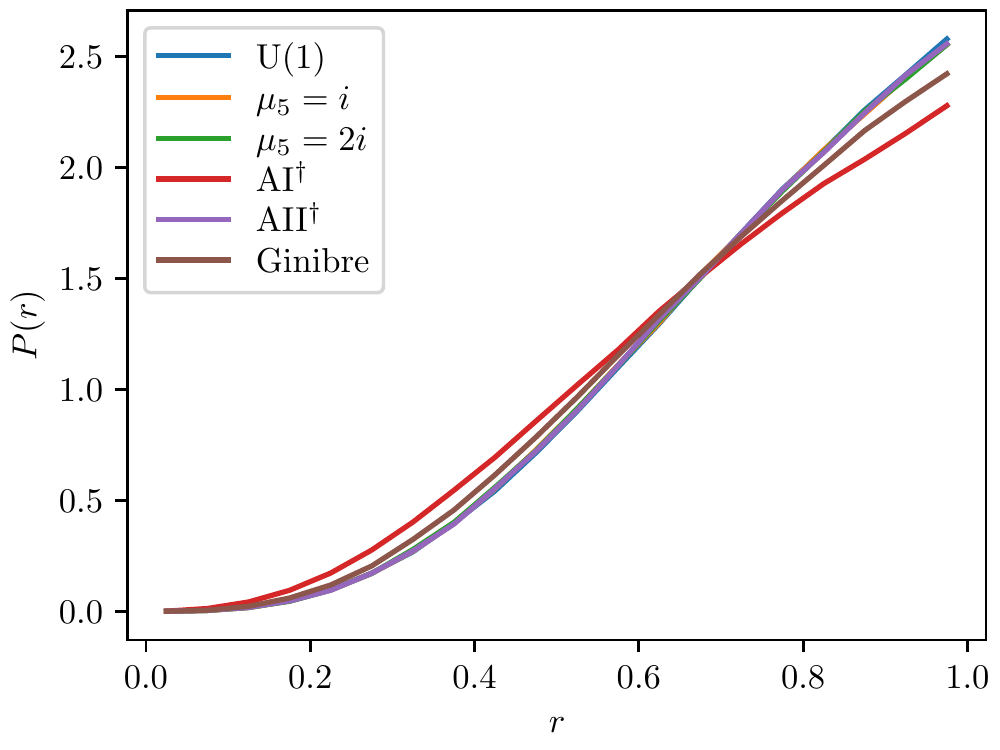}
    & \includegraphics[width=.484\columnwidth]{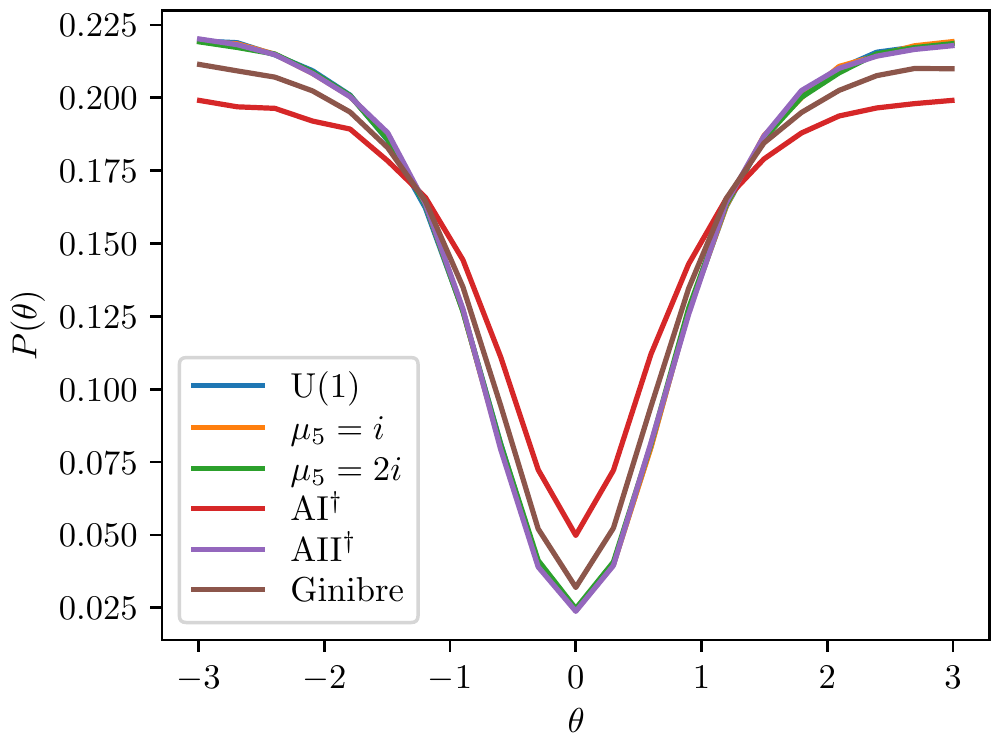} \\
    \includegraphics[width=.484\columnwidth]{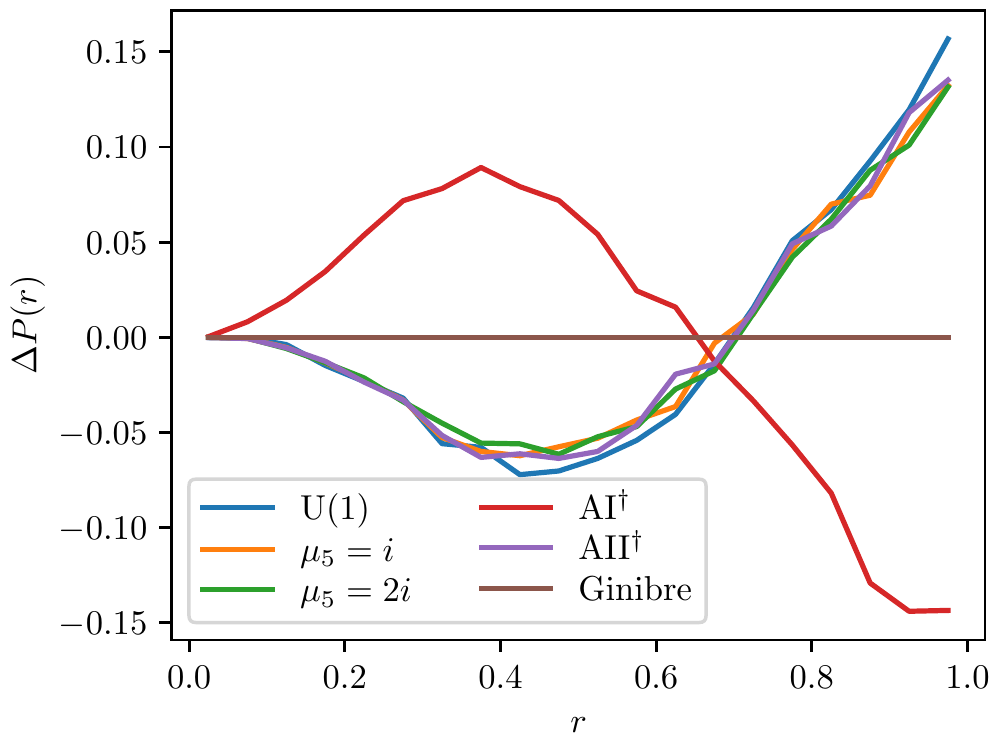}
    & \includegraphics[width=.484\columnwidth]{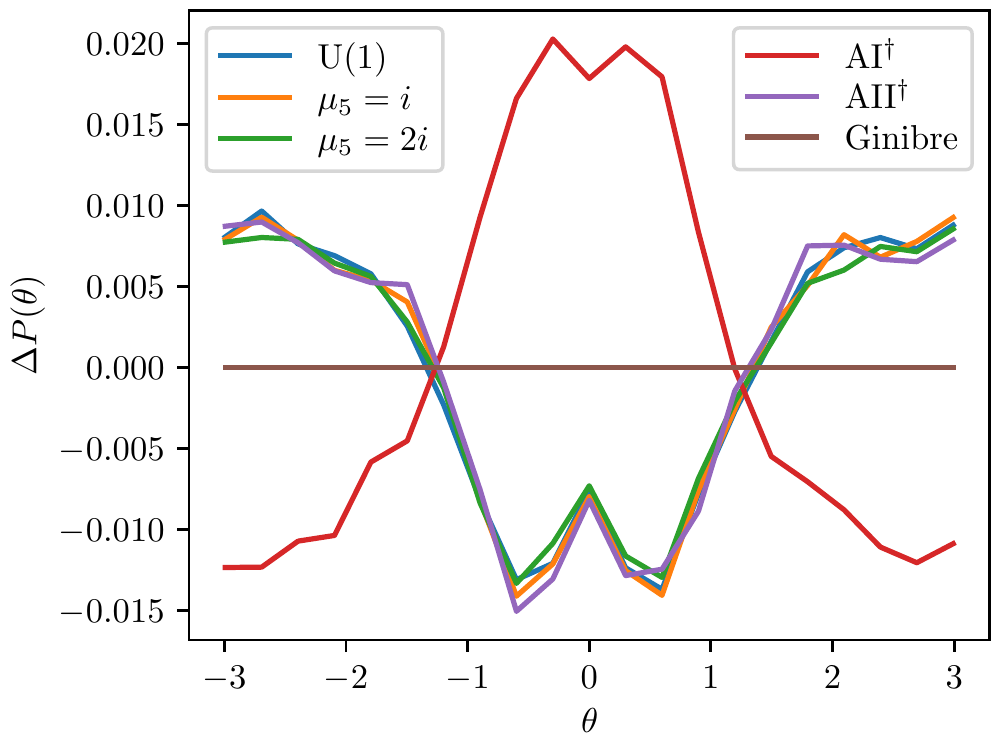}
  \end{tabular}
  \caption{Results for $P(r)$ (left) and $P(\theta)$ (right) from staggered lattice data for SU(2) fundamental and from RMT obtained by integrating the distributions $P(z)$ in Fig.~\ref{fig:2d_fund} over $\theta$ and $r$, respectively. The histograms have been plotted as lines for better visibility (horizontal values correspond to bin centers). The four curves for the \AIId\ ensemble and the lattice data are essentially on top of one another. In the bottom row we plot the same data as in the top row, but relative to the Ginibre ensemble for better visibility.}
  \label{fig:1d_fund}
\end{figure}

\begin{figure}
  \centering
  \includegraphics[width=\columnwidth]{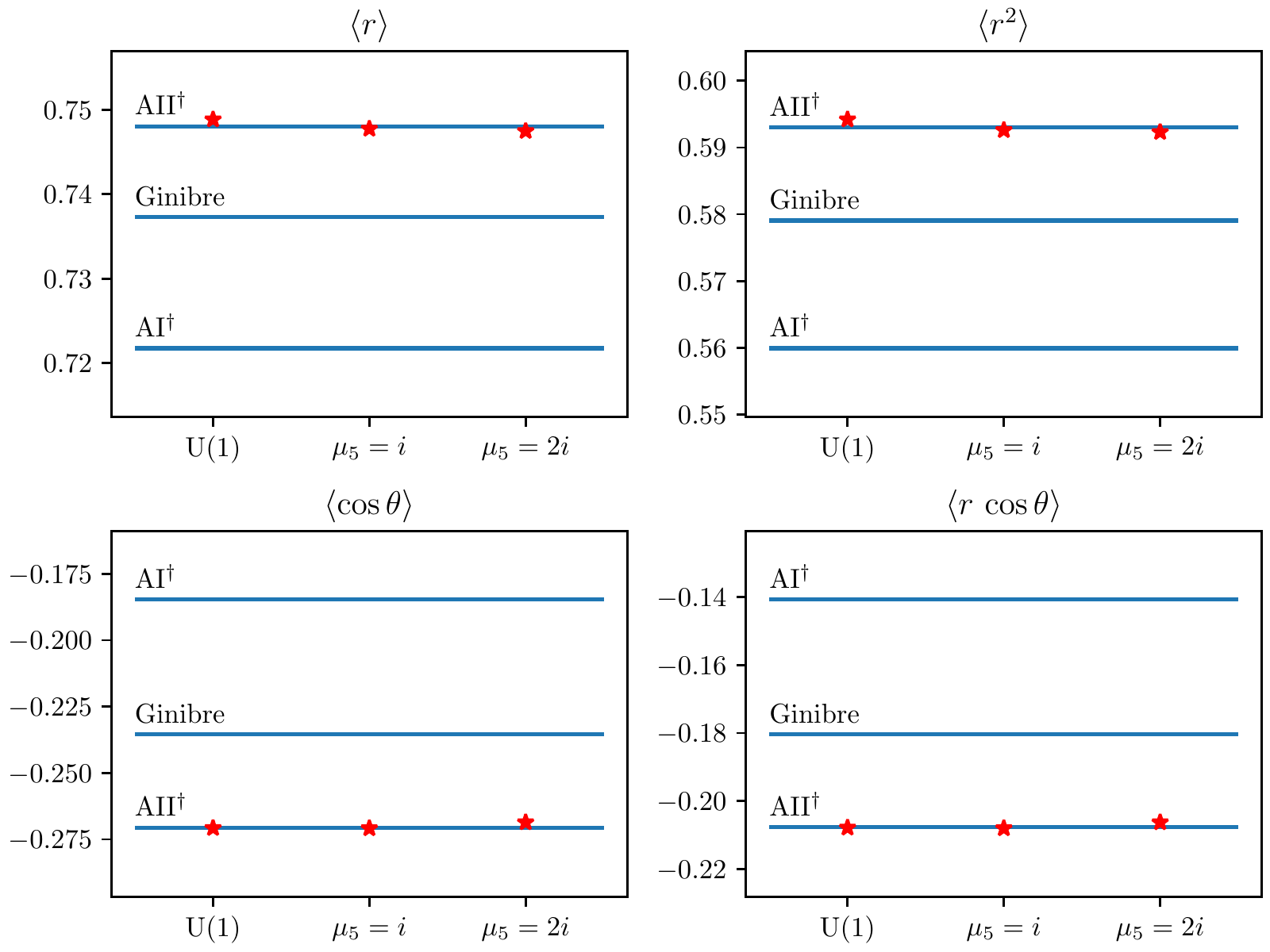}
  \caption{Moments of $P(z)$ for staggered SU(2) fundamental. The red data points are lattice data, the blue lines RMT predictions. Statistical errors are smaller than the symbols or the line thickness. We interpret the small deviations of the data from the \AIId\ lines to be boundary effects, see the comment after Eq.~\eqref{eq:zk}.}
  \label{fig:moments_fund}
\end{figure}

\begin{figure}
  \centering
  \includegraphics[width=.7\columnwidth]{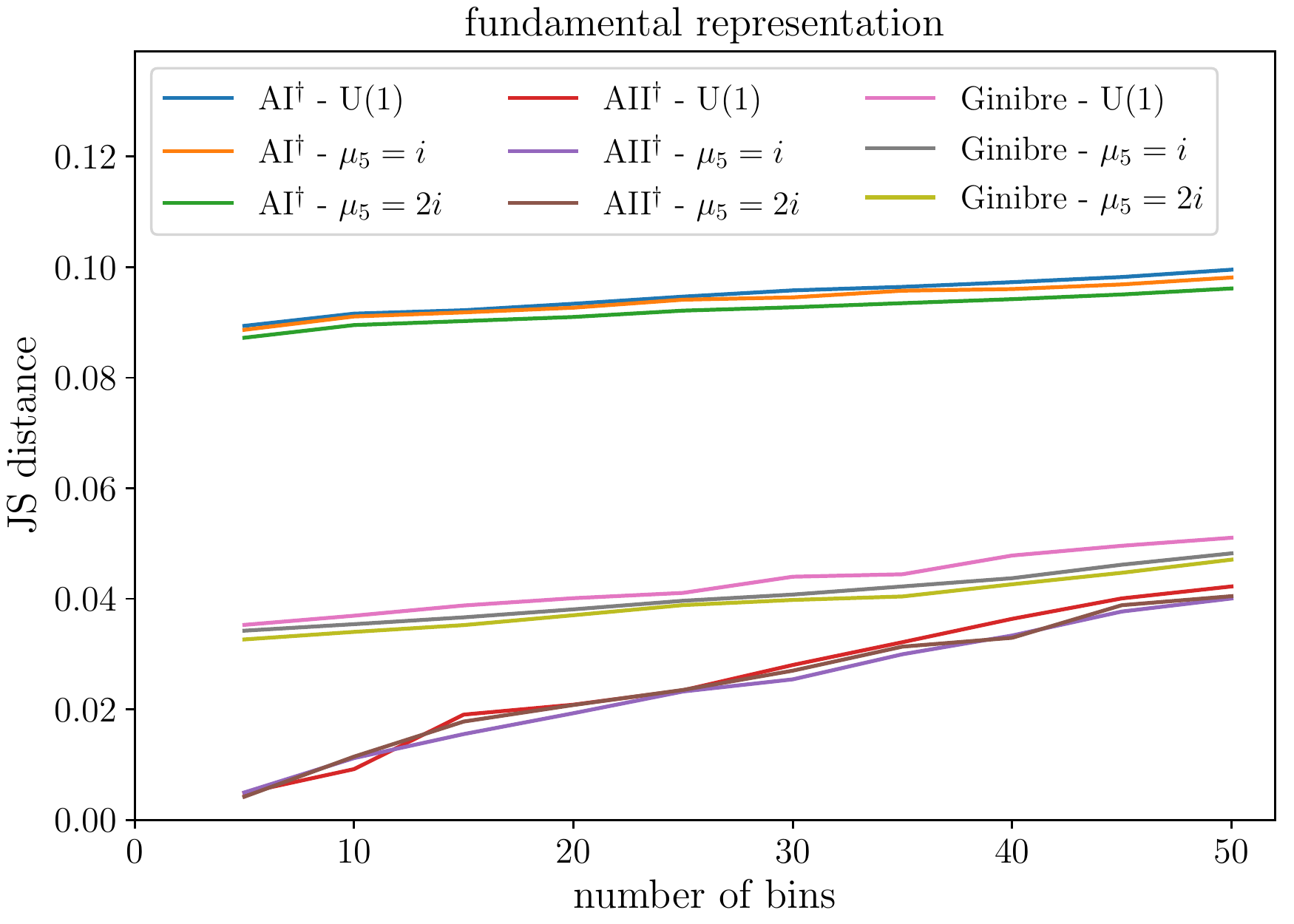}
  \caption{Jensen-Shannon distance between the distributions $P(z)$ for SU(2) fundamental in Fig.~\ref{fig:2d_fund}, computed in polar coordinates as a function of the number of bins for $r$ and $\theta$.}
  \label{fig:js_fund}
\end{figure}

Our numerical results for gauge group SU(2) and fermions in the fundamental representation are summarized in Figs.~\ref{fig:2d_fund} through \ref{fig:js_fund}. Most of the details are given in the figure captions, and therefore we only summarize our observations here. From the heatmap plots in Fig.~\ref{fig:2d_fund} we cannot draw unambiguous conclusions, but we already see that the lattice data for $P(z)$ look roughly equal for all three cases and agree with the \AIId\ class, while the \AId\ and Ginibre classes look different. From Figs.~\ref{fig:1d_fund} and \ref{fig:moments_fund} it becomes quite clear that the lattice data correspond to class \AIId. An interesting observation in Fig.~\ref{fig:1d_fund} (and Fig.~\ref{fig:1d_adj} below) is that the curves for all three ensembles appear to intersect at the same points. In the absence of analytical results we do not know whether this is actually true, but perhaps this can be shown in future work.

We also computed the Jensen-Shannon distance between pairs of the two-dimensional distributions shown in Fig.~\ref{fig:2d_fund}, comparing lattice results to RMT predictions as a function of the number of bins. The results are displayed in Fig.~\ref{fig:js_fund}. The Jensen-Shannon distance is an information-theoretic measure for how similar two probability distributions are. The fact that the distance extrapolates to zero for RMT class \AIId\ strongly supports the agreement of the data with that universality class.

\subsection{Results for SU(2) adjoint}
\label{sec:adj}

\begin{figure}
  \centering
  \begin{tabular}{cc}
    \includegraphics[width=.484\columnwidth]{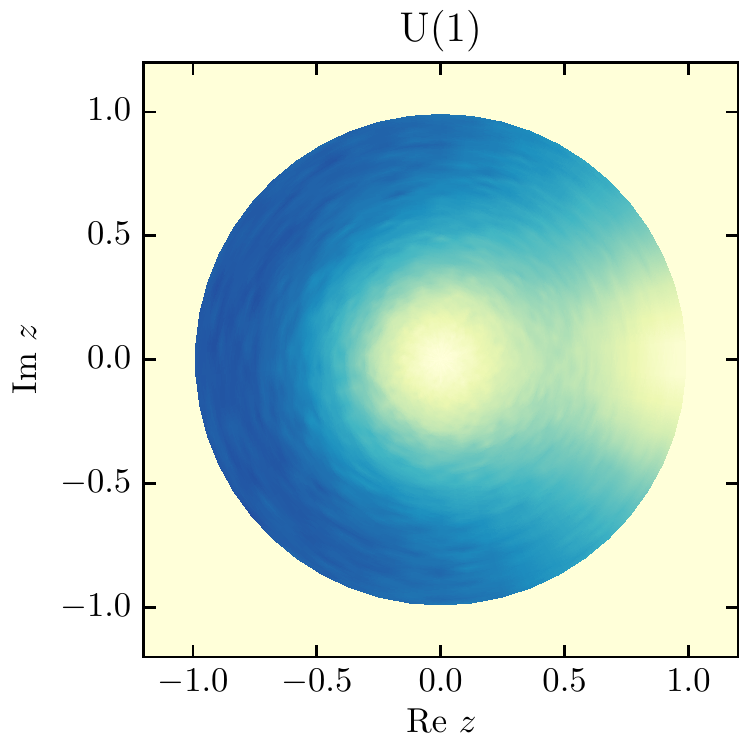}
    & \includegraphics[width=.484\columnwidth]{Pz_A1d} \\
    \includegraphics[width=.484\columnwidth]{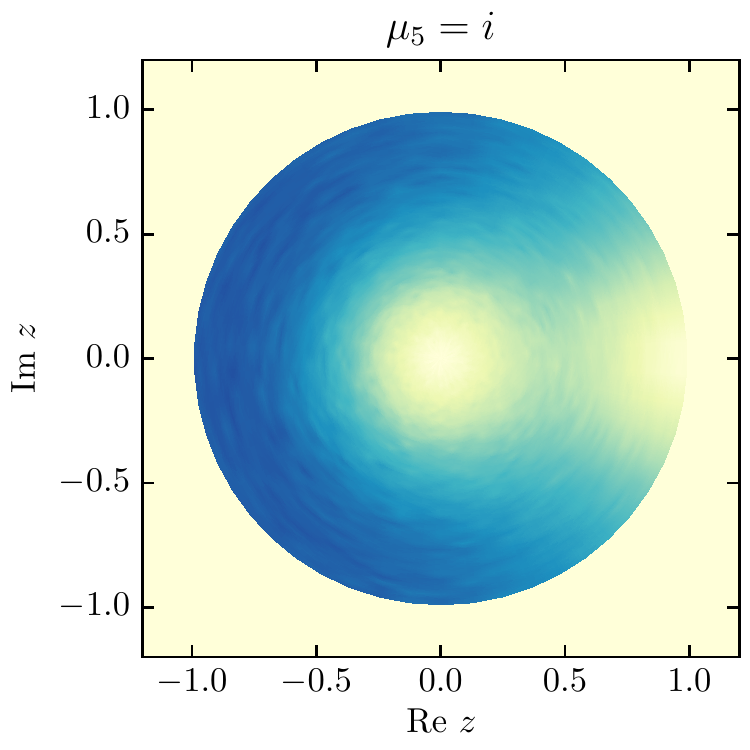}
    & \includegraphics[width=.484\columnwidth]{Pz_A2d} \\
    \includegraphics[width=.484\columnwidth]{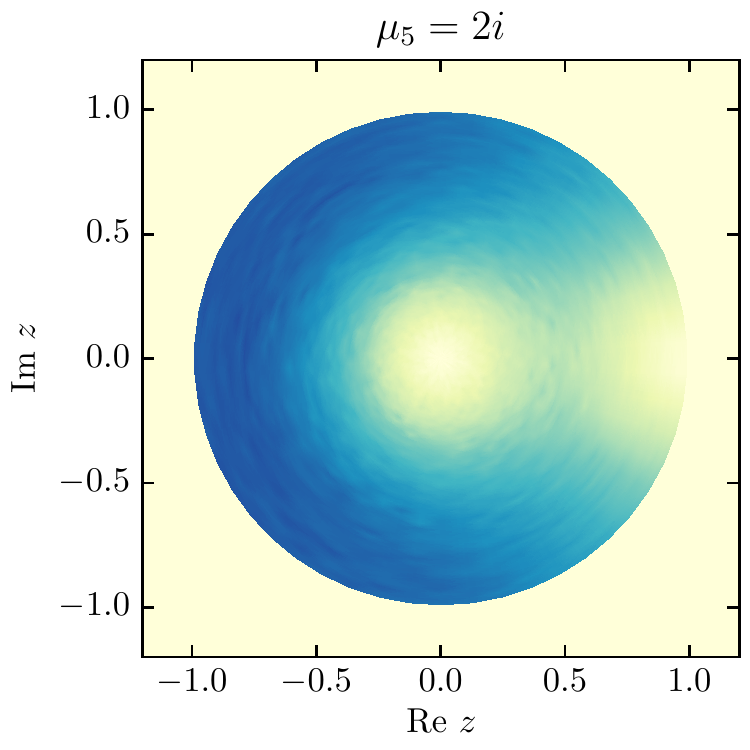} 
    & \includegraphics[width=.484\columnwidth]{Pz_Ginibre}
  \end{tabular}
  \caption{Same as Fig.~\ref{fig:2d_fund} but for the adjoint representation.}
  \label{fig:2d_adj}
\end{figure}

\begin{figure}
  \centering
  \begin{tabular}{cc}
    \includegraphics[width=.484\columnwidth]{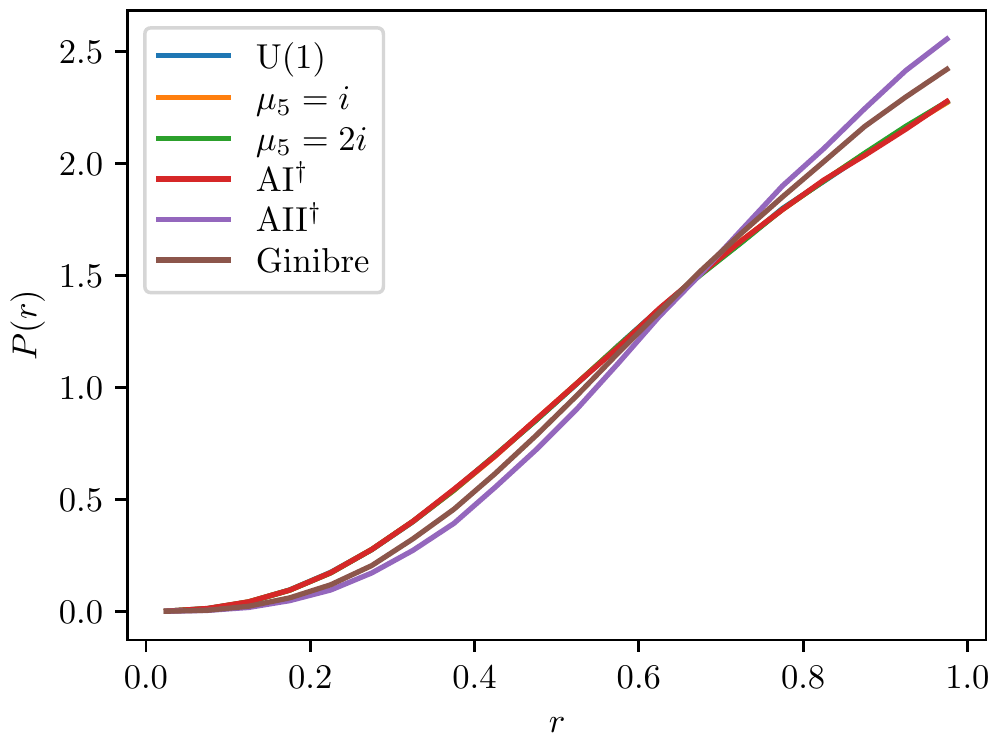}
    & \includegraphics[width=.484\columnwidth]{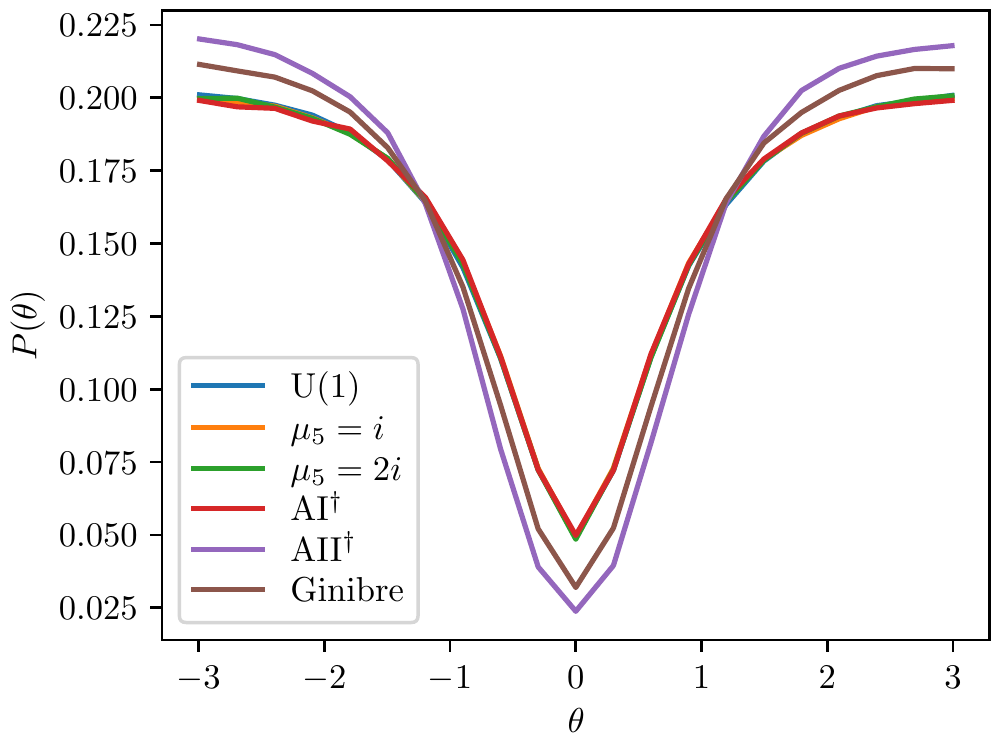} \\
    \includegraphics[width=.484\columnwidth]{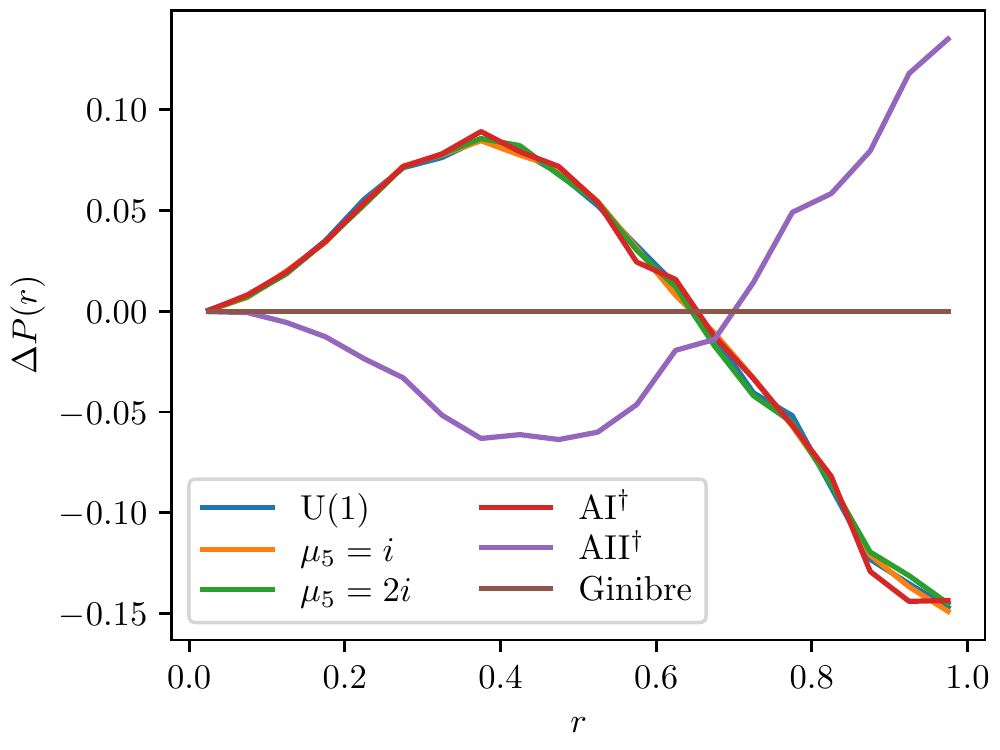}
    & \includegraphics[width=.484\columnwidth]{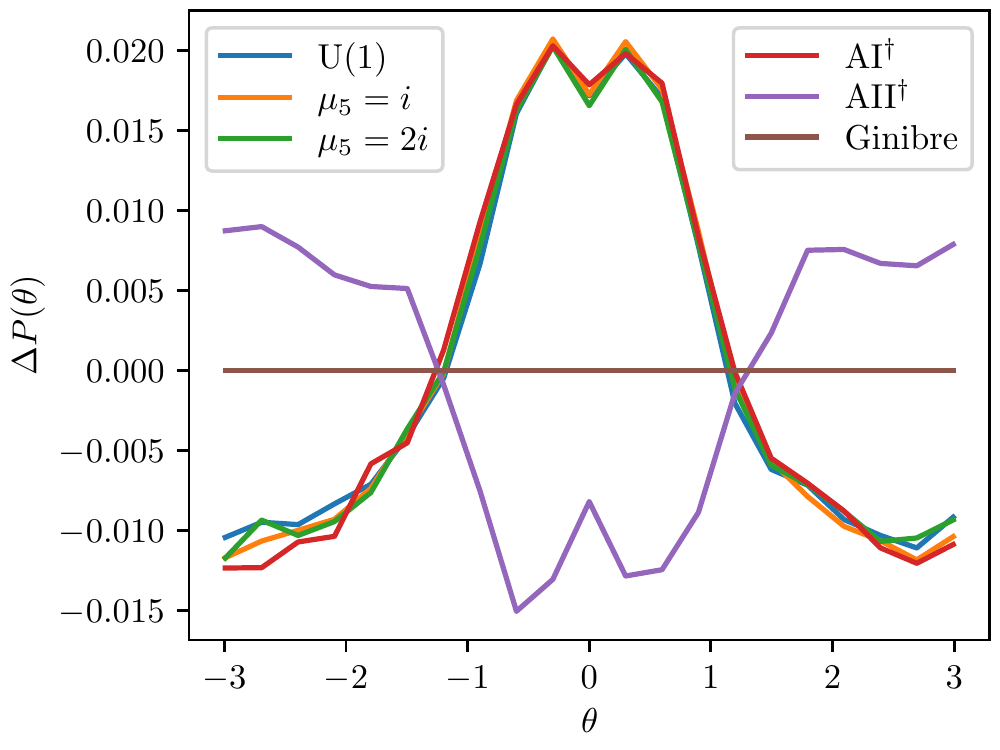}
  \end{tabular}
  \caption{Same as Fig.~\ref{fig:1d_fund} but for the adjoint representation.}
  \label{fig:1d_adj}
\end{figure}

\begin{figure}
  \centering
  \includegraphics[width=\columnwidth]{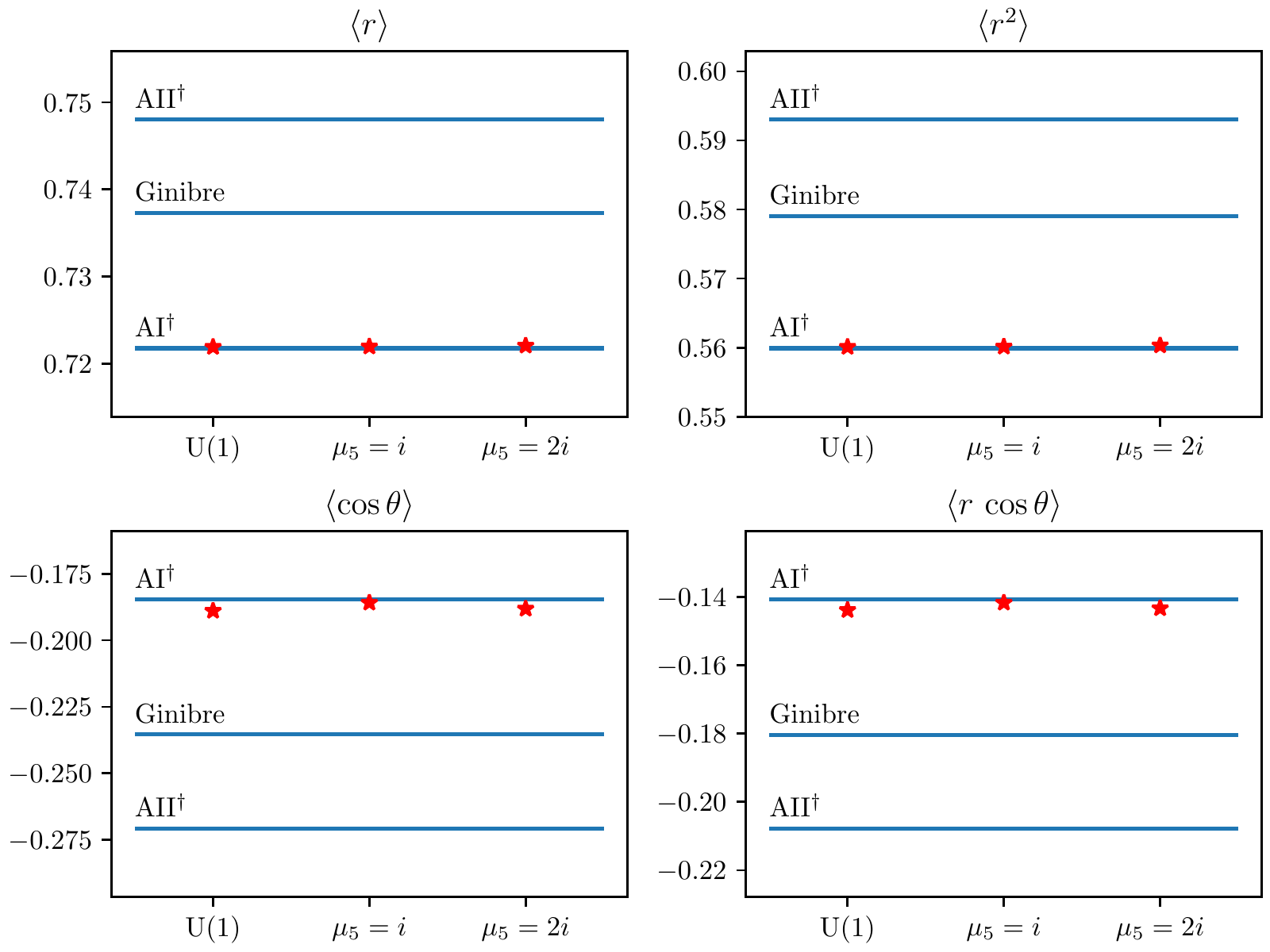}
  \caption{Same as Fig.~\ref{fig:moments_fund} but for the adjoint representation.}
  \label{fig:moments_adj}
\end{figure}

\begin{figure}
  \centering
  \includegraphics[width=.7\columnwidth]{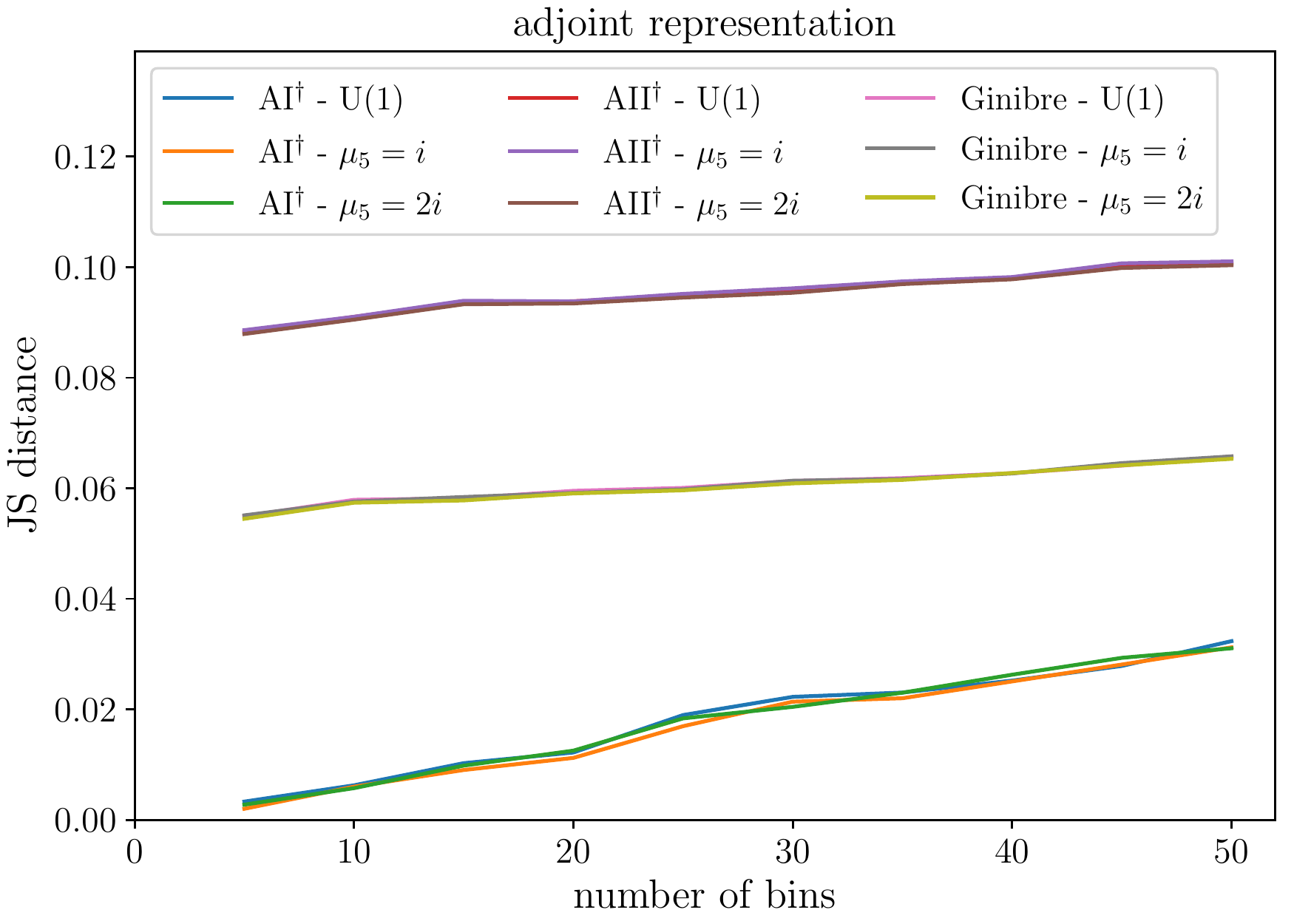}
  \caption{Same as Fig.~\ref{fig:js_fund} but for the adjoint representation.}
  \label{fig:js_adj}
\end{figure}

Our numerical results for gauge group SU(2) and fermions in the adjoint representation are summarized in Figs.~\ref{fig:2d_adj} through \ref{fig:js_adj}. The conclusions are completely analogous to those in Sec.~\ref{sec:fund}, except that we now observe agreement with RMT class \AId.

%% file: conclusions.tex
\section{Conclusions and outlook}
\label{sec:conclusions}

We have shown that the two nonstandard universality classes \AId\ and \AIId\ of non-Hermitian RMT are realized in the bulk spectral correlations of the Dirac operator in \SU(2) gauge theory coupled to a chiral \U(1) gauge field or an imaginary chiral chemical
potential. In the continuum, we find class \AId\ for pseudoreal representations of \SU(2) and class \AIId\ for real representations. We have established the corresponding chiral RMTs and verified our predictions in lattice simulations, where the above correspondence between representation and universality class is reversed for the staggered Dirac operator. We have also derived spectral sum rules that are very useful to check the numerical eigenvalues.

There are several avenues for future work. Analytically, it would be interesting to study the integral repesentation \eqref{eq:ZK} in the general case of nonequal masses. Also, one should try to derive exact RMT result for the spectral correlations, which we have generated numerically so far. On the lattice side, one could study the bulk spectral correlations in the deconfined phase, for which the same universality classes are expected. One could also focus on the local spectral correlations of the near-zero eigenvalues, for which the chiral symmetry is relevant.

\acknowledgments
This work was supported in part by DFG grant SFB TRR 55.


%% file: appendix.tex
\appendix

\section{Constant chiral gauge fields}
\label{app:axial}

Spatially inhomogeneous bilinear condensates arise in QCD and QCD-like theories under various conditions \cite{Casalbuoni:2003wh,Nakano:2004cd,Nickel:2009wj,Basar:2010zd,Buballa:2014tba,Hidaka:2015xza,Brauner:2019rjg}. To measure a modulated chiral condensate one needs to add a source term to the action,
\begin{align}
  \mathcal{O}_{\q}
  & \equiv \int \rmd \x \ckakko{\cos (2\q\cdot \x) \langle \bar\psi\psi\rangle 
		+ \sin (2\q\cdot \x) \langle\bar\psi i \gamma_5 \psi\rangle}
  \notag \\
  & = \int \rmd \x\; \langle\bar\psi \rme^{2i\gamma_5 \q\cdot \x} \psi\rangle\,,
    \label{eq:Oq}
\end{align}
which is the Fourier component of the chiral condensate with a wave vector $\q$. Its nonzero value signals spontaneous breaking of chiral, translational and rotational symmetries. The chiral limit is assumed here. By performing a space-dependent chiral rotation $\psi\to \rme^{-i\gamma_5 \q\cdot \x}\psi$ the exponential factor in \eqref{eq:Oq} may be absorbed, while a new term arises in the Lagrangian,
\begin{align}
  \bar\psi D \psi \to \bar\psi (D-i\gamma_5 \gamma_k q_k) \psi\,,
\end{align}
which has the same form as the chiral vector field $B_\nu$ in \eqref{eq:Ddefn}. The accumulation of near-zero eigenvalues of the modified Dirac operator $D-i\gamma_5 \gamma_k q_k$ is linked to the formation of modulated condensates. 

\section{Twofold degeneracy}
\label{app:twofold}

It is well-known \cite{Hands:1990wc} that the eigenvalues of the staggered Dirac operator in the fundamental representation of SU(2) are doubly degenerate since SU(2) is pseudoreal. We briefly show that this is also true in the presence of the chiral U(1) field $\theta_\mu(x)$ and the chiral chemical potential $\mu_5\in\CC$, for which $D$ is no longer anti-Hermitian. Our arguments for this non-Hermitian Kramers degeneracy closely parallel those in \cite{Kawabata:2018gjv}.

For greater generality we formally include both $\theta_\mu(x)$ and $\mu_5$ in the staggered operator. We first consider the massless case and discuss the mass term at the end of this section. We split the massless operator in the form
\begin{align}
  \label{eq:Dsplit}
  D = \frac12 D_1 + \frac12\mu_5D_5\,,
\end{align}
where $D_1/2$ and $D_5$ are given in Eqs.~\eqref{eq:U1} and \eqref{eq:D5}, respectively. Both operators satisfy \eqref{eq:2453ds}, hence so does $D$.

Now let $\ket\psi$ be a right eigenvector of $D$ with eigenvalue $\lambda$ and $\bra\phi$ be the corresponding left eigenvector with the same eigenvalue, i.e., we have
\begin{align}
  \label{eq:biorthogonal}
  D\ket\psi=\lambda\ket\psi\,,\quad\bra\phi D=\bra\phi\lambda\,,\quad\scp\phi\psi=1\,.
\end{align}
Taking the transpose of the second equation and using \eqref{eq:2453ds} we find
$D(\tau_2\ket{\phi^*})=-\lambda(\tau_2\ket{\phi^*})$. Using $\{D,\eps\}=0$ then yields
\begin{align}
  D(\eps\tau_2\ket{\phi^*})=\lambda(\eps\tau_2\ket{\phi^*})\,.
\end{align}
Hence $\lambda$ is doubly degenerate with eigenvectors $\ket\psi$ and $\eps\tau_2\ket{\phi^*}$, provided that these two vectors are linearly independent. To show this we use $(\eps\tau_2)^\T=-\eps\tau_2$ and obtain
\begin{align}
  \me\phi{\eps\tau_2}{\phi^*}=\me\phi{\eps\tau_2}{\phi^*}^\T=-\me\phi{\eps\tau_2}{\phi^*}\,,
\end{align}
i.e., we have $\me\phi{\eps\tau_2}{\phi^*}=0$. Together with the third equation in \eqref{eq:biorthogonal} this implies that 
$\ket\psi$ and $\eps\tau_2\ket{\phi^*}$ are biorthogonal and hence linearly independent.

Adding a mass term $m\1$ to $D$ simply shifts all eigenvalues by $m$, which preserves the twofold degeneracy.

\section{Derivation of spectral sum rules}
\label{app:sumrules}

In this section we give a brief derivation of the sum rules for the squared staggered Dirac operator. Again, for greater generality we include both the chiral U(1) field $\theta_\mu(x)$ and the chiral chemical potential $\mu_5$ and now also a mass term in the operator. The special cases considered in the main text, i.e., the massless staggered operator with either a chiral U(1) field or nonzero $\mu_5$, are obtained from the general sum rule \eqref{eq:sum_combined} by setting $m=0$ and either $\mu_5=0$ or $\theta_\mu(x)=1$.

We split the massive operator $D_m$ into three parts,
\begin{align}
  D_m = m\1 + \frac12 D_1 + \frac12\mu_5D_5\,,
\end{align}
where the unit matrix has dimension $\nrep V$ and the operators $D_1/2$ and $D_5$ are given in Eqs.~\eqref{eq:U1} and \eqref{eq:D5}. We trivially have $\Tr D_1=0$ and $\Tr D_5=0$ because the diagonal elements of these operators are zero. We also have
\begin{align}
  &\Tr D_1D_5=\sum_{xy}\Tr (D_1)_{xy}(D_5)_{yx}\notag\\
  &\;\;=\sum_{xy\mu}\eta_\mu(x)\Tr\left[U_\mu(x)\theta_\mu(x)\delta_{x+\mu,y}-U_\mu^\dagger(y)\theta_\mu(y)\delta_{x,y+\mu}\right]\notag\\
      &\qquad\qquad\times s(y)\bigl[\bar U_\delta(y)\delta_{y+\delta,x}+\bar U_\delta^\dagger(x)\delta_{y,x+\delta}\bigr]\notag\\
  &\;\;=0
\end{align}
since after the summation over $y$ we end up with Kronecker deltas of the form $\delta_{x\pm\mu,x+\delta}$, which are always zero because $\delta\ne\pm\mu$.\footnote{Here and below we assume that the lattice has more than two sites in every direction because otherwise some Kronecker deltas could be nonzero due to the toric nature of the lattice.}
Hence
\begin{align}
  \label{eq:D3}
  \Tr D_m^2=m^2\nrep V+\frac14\Tr D_1^2+\frac14\mu_5^2\Tr D_5^2\,.
\end{align}
We now derive the last two terms. First,
\begin{align}
  &\Tr D_1^2
  =\sum_{xy\mu\nu}\eta_\mu(x)\eta_\nu(y)\notag\\
  &\qquad\qquad\times\Tr\left[U_\mu(x)\theta_\mu(x)\delta_{x+\mu,y}-U_\mu^\dagger(y)\theta_\mu(y)\delta_{x,y+\mu}\right]\notag\\
  &\qquad\qquad\quad\:\times\left[U_\nu(y)\theta_\nu(y)\delta_{y+\nu,x}-U_\nu^\dagger(x)\theta_\nu(x)\delta_{y,x+\nu}\right]\notag\\
  &=\sum_{x\mu}\eta_\mu(x)\Bigl[-\eta_\mu(x+\mu)\theta_\mu^2(x)\Tr U_\mu(x)U_\mu^\dagger(x)\notag\\
  &\qquad\quad-\eta_\mu(x-\mu)\theta_\mu^2(x-\mu)\Tr U_\mu^\dagger(x-\mu)U_\mu(x-\mu)\Bigr]\notag\\
  &=-8\nrep V\ev{\theta_\mu^2(x)}_{x\mu},
\end{align}
where we have used $\eta_\mu(x)\eta_\mu(x\pm\mu)=1$ and the fact that $U$ is unitary. In the last line, $\ev{\cdots}_{x\mu}$ denotes an average over all links of the lattice. Furthermore,
\begin{align}
  \Tr D_5^2&=\sum_{xy}s(x)s(y)\Tr\bigl[\bar U_\delta(x)\delta_{x+\delta,y}+\bar U_\delta^\dagger(y)\delta_{x,y+\delta}\bigr]\notag\\
  &\qquad\qquad\qquad\;\;\,\times\bigl[\bar U_\delta(y)\delta_{y+\delta,x}+\bar U_\delta^\dagger(x)\delta_{y,x+\delta}\bigr]\notag\\
  &=\sum_xs(x)\bigl[s(x+\delta)\Tr \bar U_\delta(x)\bar U_\delta^\dagger(x)\notag\\
  &\qquad\qquad+s(x-\delta)\Tr \bar U_\delta^\dagger(x-\delta)\bar U_\delta(x-\delta)\bigr]\notag\\
  &=-2V\,\bigl\langle\Tr \bar U_\delta(x)\bar U_\delta^\dagger(x)\bigr\rangle_x\,,
\end{align}
where we have used $s(x)s(x\pm\delta)=-1$ and $\ev{\cdots}_{x}$ denotes an average over all sites of the lattice. Note that $\bar U_\delta(x)$ is not unitary.
From \eqref{eq:D3} we obtain
\begin{align}
  \label{eq:sum_combined}
  &\Tr D_m^2=\sum_n\lambda_n^2\\
  &=V\Big[\nrep\left(m^2\!-\!2\ev{\theta_\mu^2(x)}_{x\mu}\right)
    \!-\!\frac12\mu_5^2\bigl\langle\Tr\bar U_\delta(x)\bar U_\delta^\dagger(x)\bigr\rangle_x\Big]\,.
    \notag
\end{align}
In our numerical simulations $\mu_5$ is purely imaginary.
